# Direct readout of excited state lifetimes in chlorin chromophores under electronic strong coupling

Alexander M. McKillop[1], Liying Chen[1], Ashley P. Fidler[1,†], and Marissa L. Weichman[1,*]

[1]Department of Chemistry, Princeton University, Princeton, New Jersey, 08544, United States

**ABSTRACT:** The mechanisms governing molecular photophysics under electronic strong coupling (ESC) remain elusive to date. Here, we use ultrafast pump-probe spectroscopy to study the nonradiative excited state relaxation dynamics of chlorin e6 trimethyl ester (Ce6T) under strong coupling of its transition from the electronic ground state to the $Q_y$ band. We use dichroic Fabry-Pérot cavities to provide a transparent spectral window in which we can directly track the excited state population following optical pumping of either the strongly-coupled $Q_y$ band or the higher-lying B band. This scheme circumvents many of the optical artifacts inherent in ultrafast cavity measurements and allows for facile comparison of strongly-coupled measurements with extracavity controls. We observe no significant changes in excited state lifetimes for any optical pumping schemes or cavity coupling conditions considered herein. These results suggest that Ce6T exhibits identical photophysics under ESC and in free space, presenting a new data point for benchmarking emerging theories for cavity photochemistry.

## 1. INTRODUCTION

Structure-function relationships govern the photophysical properties of molecules, permitting the tunability of excited state energies, lifetimes, and relaxation pathways.[1–4] However, the parameter space is vast and one must typically synthesize and screen a library of molecular candidates in order to optimize a desired photochemical process.[5–7] Over the past two decades, molecular polaritons have emerged as a new tool to potentially modulate the photophysics[7–22] and photochemistry[23–26] of organic molecules without structural modification.[7,16,27,28] Polaritons are hybrid light-matter states that arise from strong coupling of the bright optical transition of an ensemble of molecules to a confined electric field, often engineered within an optical cavity.[27,29–36] A comprehensive understanding of how molecules behave under cavity coupling of electronic transitions is still lacking, particularly with regard to their ultrafast dynamics. Here, we directly probe the excited state dynamics of chlorin e6 trimethyl ester (Ce6T) (Fig. 1a) under electronic strong coupling (ESC) in dichroic Fabry-Pérot (FP) cavities.

Polaritons manifest when the rate of energy exchange between a cavity field and intracavity molecular dipoles surpasses that of their individual decay mechanisms, splitting the coupled cavity resonance into two new modes separated in energy by the Rabi splitting ($\hbar\Omega_R$).[29,30,37] Polariton formation can be described with collective cavity quantum electrodynamics (cQED) frameworks like the Tavis-Cummings model,[38] or using classical optics methods like the transfer matrix method (TMM) and Fabry-Pérot cavity expressions.[39–47] Figure 1b depicts a simplified view of the Tavis-Cummings model where $N$ molecular dipoles couple to a single, discrete cavity mode, forming upper and lower polariton states (UP, LP) in addition to $N-1$ dark states that are purely molecular in character.[29,30,35,37]

Both the cQED and classical approaches reproduce the scaling of the Rabi splitting with the molecular transition dipole and the square-root of the concentration of intracavity molecules ($\sqrt{N/V}$), where $N$ is the number of coupled molecules and $V$ is the cavity mode volume.[48–50]

The electronic transitions of organic molecules were first cavity-coupled to form exciton-polaritons by Lidzey et al. in 1998.[51] It has since been an open question whether molecules prepared in electronic polariton states feature distinct photochemical behavior as compared to excited molecules in free space. Ebbesen and coworkers began laying the foundations to address this question in the 2010s, focusing initially on ESC of molecular photoswitches.[23,52] An influential 2012 report from Hutchison et al.[23] observed slowing of the unimolecular isomerization of spiropyran to merocyanine in a thin film under ESC of the $S_0 \rightarrow S_1$ transition in merocyanine using continuous wave (cw) irradiation. This initial work spurred major theoretical efforts to understand photochemistry under ESC.[17,37,53–66] A significant body of emerging theory predicts that ESC can reshape the coupled excited state potential energy surface, creating a local minimum that traps population in the Franck-Condon region.[54,55,57] Even with this tantalizing hypothesis, only a handful of experimental efforts have subsequently reported on photochemistry under ESC. Hutchison's results have been reproduced by other groups[25,26,67] and extended to other photoswitches,[15,25] though questions remain about the reproducibility of these findings.[67–70] In parallel, other groups have reported suppressed photodegradation rates for dyes under ESC.[10,71,72] Dutta et al.[73] observed slowed excited state intramolecular proton transfer under ESC and explained this result using optical filtering arguments, without invoking cavity-alteration of excited state surfaces.

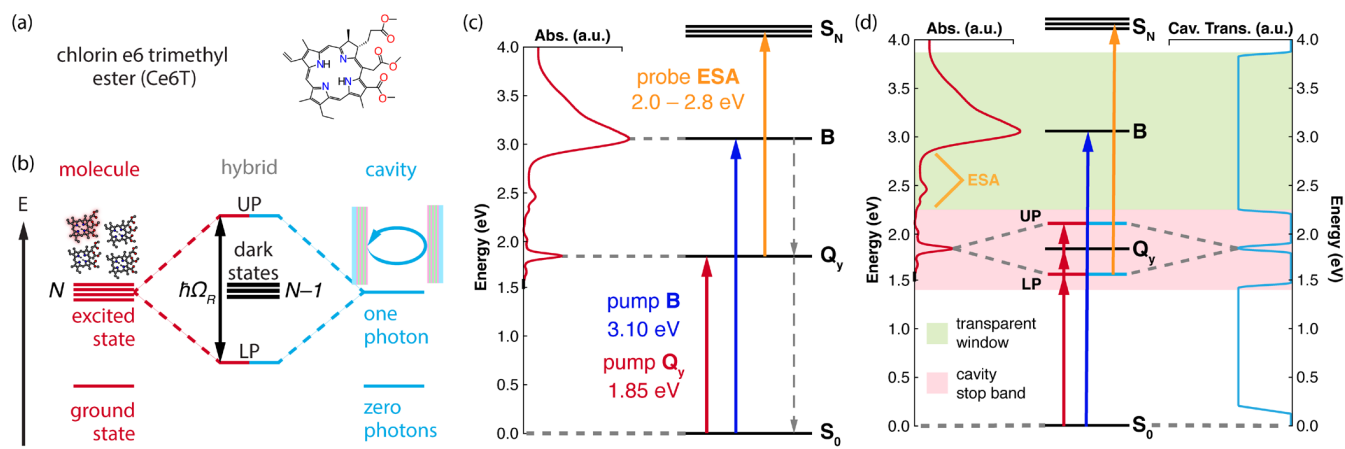

**Figure 1.** (a) Chemical structure of chlorin e6 trimethyl ester (Ce6T). (b) Collective strong light-matter coupling in a Fabry-Pérot cavity. Within the Tavis-Cummings model, $N$ molecules couple to the quantized excitation of a single cavity mode, yielding upper and lower polariton (UP, LP) states as well as $N - 1$ molecular dark states. (c) Simplified energy diagram of Ce6T, with the absorption spectrum of a Ce6T thin film plotted on the left in red. The $Q_y$ and B bands lie 1.86 eV (668 nm) and 3.07 eV (404 nm) above the $S_0$ ground state, respectively. Optical pumping can be performed for either the $Q_y$ band (red arrow) or B band (blue arrow); the latter undergoes rapid nonradiative relaxation to $Q_y$. The population of $Q_y$ can be tracked via a broad excited state absorption (ESA, yellow arrow) that appears in the range of 2.0–2.8 eV (440–620 nm). (d) Cartoon of experiments to probe the photophysics of Ce6T under electronic strong coupling. The transmission spectrum of an idealized dichroic cavity is plotted on the right in light blue, with reflective stop band highlighted in pink. The supported cavity mode is coupled to the $S_0 \rightarrow Q_y$ transition of Ce6T. Transparency of the cavity at higher energies (pale green area) allows for direct spectroscopic access to the ESA feature (yellow arrow) following optical pumping of the strongly-coupled $Q_y$ manifold (red arrows) or the B band (blue arrow)

Most studies of photochemistry under ESC rely on steady-state readouts acquired under cw illumination on timescales of minutes to hours that far exceed the intrinsic dynamical time constants. On the other hand, there is a growing body of ultrafast work on molecules under ESC using pump-probe[8,12–14,17,19,74–78] and multidimensional[79,80] spectroscopy. Some of these efforts report modifications to excited state lifetimes[8,81,12,13,19] under ESC. Ultrafast spectroscopy provides a powerful means to investigate fundamental molecular behavior under ESC, but results can be difficult to interpret as excited state signals are filtered through optical cavity transmission windows that counter-intuitively distort under optical pumping of the intracavity material.[47,76,82–85] Transient spectra acquired under strong coupling therefore contain signatures arising from both the intrinsic dynamics of intracavity molecules and the response of the cavity,[76] making direct comparison to free-space measurements challenging. Methods for accurate extraction of intracavity molecular dynamics from nonlinear cavity spectra are still emerging.[8,47,84,86]

In light of these complexities, mechanistic understanding of the impact of polariton formation on photochemistry remains lacking. The role that excited state lifetimes play in modulating photochemical rates under ESC remains a debated topic.[17,23,25,73,87] It is also unknown how polariton photophysics depends on excitation scheme – e.g. direct optical pumping of the polariton manifold versus excitation of higher-lying states that decay and populate the polaritons indirectly. More broadly, it is unclear whether reports of altered intracavity photochemistry derive specifically from ESC or can instead be rationalized by optical artifacts, cavity-enhanced absorption, and/or heterogeneous disorder.[35,73,84,47,67–70]

Here, we examine the ultrafast photophysics and excited state lifetimes of Ce6T chromophores under ESC. Ce6T exhibits two prominent electronic bands that absorb in the visible: the $S_0 \rightarrow S_1$ transition ($Q_y$ band) at 1.86 eV (668 nm) and the $S_0 \rightarrow S_2$ transition (B band) at 3.07 eV (404 nm) (Fig. 1c).

The ultrafast dynamics following excitation of these bands in chlorins and related porphyrinoids are well studied.[88–91,19] In thin film samples of these species, the $Q_y$ band decays predominantly via relaxation to a delocalized excimer state and via nonradiative internal conversion to the ground state; radiative and intersystem crossing decay channels participate minimally.[89,91,92] Optical excitation of the $Q_y$ band produces an excited state absorption (ESA) feature that spans 2.0–2.8 eV (440–620 nm). Optical excitation of the higher-lying B band leads to rapid, sub-picosecond population relaxation to the $Q_y$ band, which then goes on to relax non-radiatively.[88,89,93–95] In addition to having well-characterized ultrafast dynamics, Ce6T is convenient for strong cavity coupling as the $Q_y$ band features both a large transition dipole and a narrow linewidth. These advantages have made both Ce6T and its non-trimethylated analogue Ce6 prime targets for studies of photophysics under ESC.[9,18,19]

Here, we place thin films of Ce6T in FP cavities constructed from dichroic distributed Bragg reflectors (DBRs). Dichroic mirrors permit strong coupling in one spectral region while maintaining sufficient transmission in another region for ultrafast pump-probe spectroscopy free from cavity interference and optical artifacts.[96,97] In particular, we engineer cavities that support an optical resonance near 1.86 eV for strong coupling the Ce6T $Q_y$ band while remaining highly transmissive from 2.2 to 2.8 eV for direct optical access to track transient ESA signatures and read out excited state lifetimes (Fig. 1d). We record pump-probe spectra following either direct optical excitation of the polaritonic $Q_y$ manifold near 1.86 eV or indirect population via optical pumping of the B band and subsequent relaxation into the polaritonic $Q_y$ manifold. Regardless of excitation scheme, we detect no statistically significant change in $Q_y$ band lifetimes under ESC. We additionally find no dependence of the ultrafast dynamics on the Rabi splitting or the coupled cavity mode order.

The chief finding of this work is that Ce6T exhibits identical excited-state photophysics under ESC and in free space. This outcome is consistent with the large separation of timescales in our system, where the excited-state lifetimes of Ce6T far exceed the cavity photon lifetime. Such behavior aligns with emerging theoretical and experimental evidence suggesting that long-lived excited states are largely insensitive to perturbation with ESC.[73,78,98] As the mechanisms of cavity photochemistry are not yet fully understood, this result represents a new data point against which to benchmark polariton theory.

## 2. EXPERIMENTAL METHODS

We use ultrafast pump-probe spectroscopy to directly quantify excited state population dynamics in Ce6T thin films embedded in dichroic FP cavities. We describe our pump-probe spectrometer in Section 2.1; a more detailed description of this instrument is available in our prior publications.[96,97] We describe linear spectroscopy techniques used to characterize samples in Section 2.2. We describe fabrication of DBR mirrors and assembly of optical FP nanocavities in Section 2.3. In Section 2.4, we describe a rastering scheme which allows us to perform ultrafast measurements over select sample areas with uniform cavity-coupling conditions.

**2.1. Ultrafast visible-pump/white-light-probe spectroscopy.** Our visible-pump, white-light-probe setup is driven by a Ti:sapphire femtosecond laser system (Astrella, Coherent) that produces 7 mJ/pulse, 60 fs pulses centered at 800 nm with a 1 kHz repetition rate. To generate tunable visible pump pulses, 2.5 mJ of the fundamental laser light is directed to an optical parametric amplifier (OPA, OPerA Solo, Light Conversion). We use the second harmonic (SH) crystal in the OPA to double the Ti:sapphire fundamental to generate pump light for experiments performed with excitation at 3.10 eV (400 nm). We use the SH-signal OPA configuration to generate pump light for excitation at 1.81 eV (684 nm), 1.90 eV (653 nm), and 1.85 eV (669 nm). We attenuate pump pulses with a variable neutral density filter (OD: 0.02–2.0, Thorlabs) such that 90 to 150 nJ pulse energies strike the sample. The 3.10 eV pump beam has a slightly elliptical profile with a diameter of ~200 μm at focus, yielding typical pump fluences between 255 and 426 μJ/cm$^2$. The pump fluences used at other excitation energies are similar. Before the sample, a mechanical chopper (MC2000, Thorlabs) blocks every other pump pulse at 500 Hz to allow for shot-to-shot subtraction. We characterize pump spectra using a fiber-coupled spectrometer (BLUE-Wave UVNb-25, StellarNet). Representative pump spectra are provided in Section S1 of the Supporting Information (SI).

In the probe arm, we direct 1.1 mJ of fundamental Ti:sapphire laser light to a 325 mm long motorized delay line (DL325, Newport) employing a single-pass geometry that yields 2.2 ns of optical delay. After the delay line, we focus the light into a 3 mm thick, flat [001] CaF$_2$ crystal (Eksma Optics) to generate our white light continuum (WLC) probe. To reduce burning, we translate the CaF$_2$ crystal continuously through a random set of Gaussian distributed points using an actuated stage. We optimize the focusing and power of 800 nm light incident on the CaF$_2$ crystal to stabilize the WLC intensity fluctuations to a variance of <4% across our probing region. At focus, the probe beam has a power of 600 nJ and a diameter of ~100 μm, yielding a typical probe fluence of ~7 mJ/cm$^2$. The probe spectrum is provided in Section S1 of the SI.

The WLC probe is aligned at normal incidence to the sample, while the pump beam is aligned to overlap the probe at the sample with a crossing angle of ~8°. In all experiments reported here, the relative polarization of pump and probe beams is set near the magic angle. We raster the sample in 100 μm increments and with 1 second dwell times using motorized delay stages (Zaber) controlled with home-written MATLAB software. After interacting with the sample, the WLC probe enters a Schmidt-Czerny-Turner grating spectrometer fitted with a charge-coupled device camera (Isoplane-320 and Blade-400B, Princeton Instruments). All transient spectra are acquired using a home-written LabVIEW program. We report transient data in differential optical density (ΔOD):

$$\Delta\text{OD} = -\log_{10}\left[\frac{I_{\text{pump-on}}}{I_{\text{pump-off}}}\right]$$

where $I_{\text{pump-off}}$ and $I_{\text{pump-on}}$ represent the intensity of probe light transmitted through the sample with the pump beam blocked and unblocked, respectively. We chirp correct all pump-probe data using a second-order polynomial over the grating spectrometer's wavelength axis. Temporal linecuts of transient pump-probe data are fit to exponential functions to extract time constants using the nonlinear least squares method in MATLAB.

**2.2. Linear spectroscopic characterization methods.** We characterize the optical properties of thin film samples, solution-phase samples, and cavity devices using various additional absorption and emission spectroscopies. We use a Cary 60 UV-Vis spectrometer to collect absorption spectra of extracavity thin film and solution-phase samples as well as transmission spectra of mirrors and cavity devices. We use a Cary 5000 UV-Vis spectrophotometer with a Universal Measurement Attachment (minimum angle of 10°) to collect reflection spectra of DBR mirrors. We use an Edinburgh Instruments FLS980 spectrometer to collect photoluminescence (PL) spectra and time-correlated single photon counting (TCSPC) data for extracavity thin film and solution-phase samples. The instrument response function (IRF) of the TCSPC system is described in Section S2 of the SI.

**2.3. Sample preparation.** To minimize spectral artifacts for pump-probe measurements, we fabricate dichroic FP cavities composed of pairs of DBR mirrors sandwiching a thin film of Ce6T in polystyrene (PS), as illustrated in Fig. 2a.

*2.3.1. DBR mirror design.* We design DBR mirrors with a target reflectivity of $R$~70% for light at 1.7−2.0 eV (620−730 nm) to enable strong coupling of the Ce6T S$_0$→Q$_y$ band transition at 1.86 eV. We additionally require the DBR transmission to be $T$>90% from 2.0−2.8 eV (440–620 nm) to provide a transparent spectral window for pump-probe measurements. We design mirrors with these constraints using TMM[41,42,45] simulations implemented in MATLAB, making use of literature refractive indices.[99–101] We arrive at an optimal design of 3.5-pair, λ/4 DBR mirrors composed of alternating layers of high-$n$ Si$_3$N$_4$ and low-$n$ SiO$_2$. These mirrors feature a simulated $R$~73% reflectivity at 1.86 eV and $T$~90% transmission from 2.2−2.8 eV (Fig. 2b).

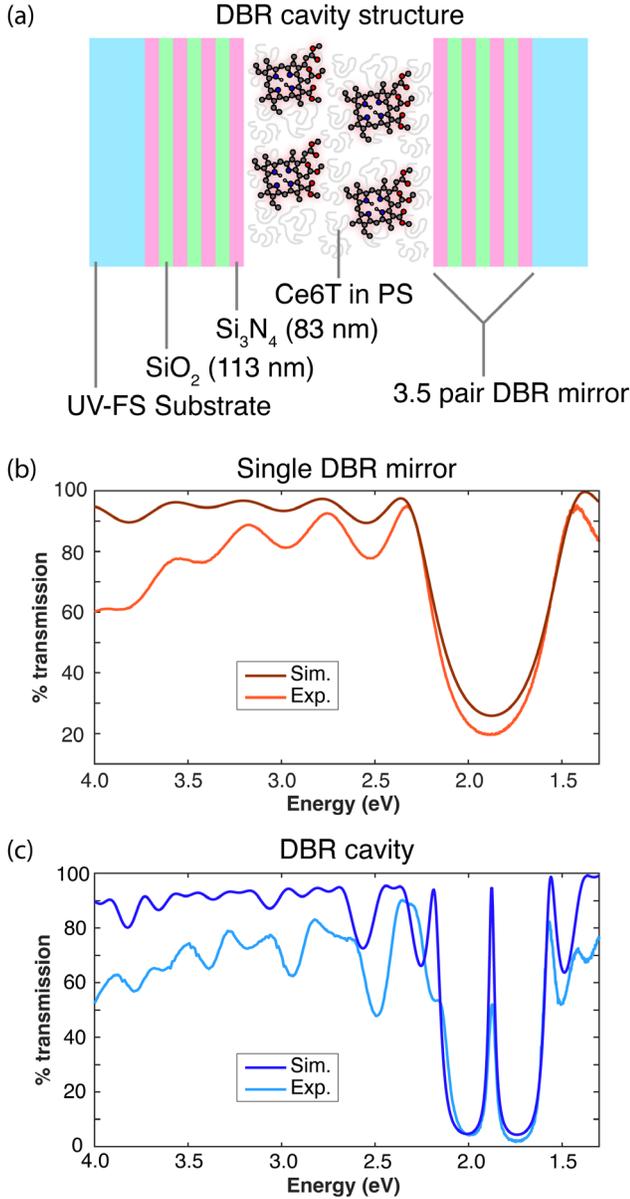

**Figure 2.** Structure and characterization of dichroic distributed Bragg reflector (DBR) mirrors and cavity assemblies. (a) Fabry-Pérot (FP) cavities are formed by sandwiching two DBR mirrors around a polystyrene (PS) film containing Ce6T molecules. Each DBR mirror is made of 3.5 pairs of high-$n$ $Si_3N_4$ and low-$n$ $SiO_2$ deposited on a UV-fused silica (UV-FS) substrate. (b) Experimental (light orange) and simulated (dark orange) transmission spectra for a single representative DBR mirror at normal incidence. (c) Experimental (light blue) and simulated (dark blue) transmission spectra for an assembled control two-mirror FP cavity containing a ~711 nm PS thin film at normal incidence, showing the formation of a cavity mode at 1.87 eV.

*2.3.2. DBR mirror fabrication.* We fabricate DBR mirrors in-house using chemical vapor deposition with an Oxford PlasmaPro 100 ICP-CVD system. Before each coating run, we deposit calibration layers of $Si_3N_4$ and $SiO_2$ onto silicon wafers and quantify the deposition rates and refractive indices of both materials using a Gaertner ellipsometer equipped with a single-wavelength 632 nm laser. With accurate measures of these parameters, we then deposit 3.5 layer pairs (4 layers of ~83 nm $Si_3N_4$ and 3 layers of ~113 nm $SiO_2$) onto 1" and ½" diameter, 3 mm thick UV-fused silica (UV-FS) optical substrates (Eksma Optics). We characterize the transmission and reflection spectra of the fabricated mirrors, finding good agreement with TMM simulations (Fig. 2b). The transmission of the fabricated mirrors does drop off at higher energies as compared to simulation likely due to interlayer losses.

*2.3.3. Sample spin coating.* We spin coat Ce6T/PS films onto bare UVFS or mirror substrates using a Specialty Coating Systems 6800 spin coater. Following previous literature,[18] our spin-coating solution contains 8 wt. % PS (MilliporeSigma, MW ~192,000) in toluene (MilliporeSigma, >99.5% ACS Reagent) with 16 wt. % Ce6T (Frontier Specialty Chemicals, >95%) in PS. We heat the PS/toluene solution at 55°C to ensure full dissolution prior to adding Ce6T.

The Ce6T/PS film thickness is important as it sets the length of the final cavity device. We target film thicknesses of 809 nm and 979 nm in two polaritonic cavities which permit, respectively, resonant normal-incidence coupling of the fourth- and fifth-order longitudinal FP cavity modes to the Ce6T $S_0 \rightarrow Q_y$ band at 1.86 eV. We hereafter refer to these two devices as Cavity 1 and Cavity 2. We prepare additional control devices (Cavities 3−6) which are detuned from resonance at normal incidence with film thicknesses between 814 nm and 955 nm. This film thickness range provides sufficient pathlength for workable signal:noise in pump-probe measurements while remaining thin enough that the resulting cavities feature a sufficiently large free spectral range (FSR) for clean cavity-coupling conditions. Designing cavities that are resonant at normal incidence is essential to permit rastering of the sample in pump-probe experiments; see Section 2.4 below.

To achieve films of the desired thickness, we construct a calibration curve for film thickness versus spin rate for each batch of Ce6T/PS/toluene solution. We prepare a series of calibrant thin films on UV-FS substrates with spin-coating rates ranging from 1200 to 2800 rpm. We dry the calibrant films in a vacuum desiccator overnight followed by placing them on a 110°C hotplate for 6 hours. We verify that this heat exposure does not degrade Ce6T by acquiring UV-Vis and pump-probe spectra after various heating times. We use a KLA Tencor P-17 profilometer to characterize the film thicknesses and construct a calibration curve. We finally spin coat the 1" DBR mirrors with Ce6T/PS at the appropriate calibrated spin rate to yield films of the desired thickness. We then dry the films at 110°C and characterize their thicknesses using profilometry to select for the films which will give resonant cavities before moving on to cavity assembly. Spin-coating runs that result in undesired film thicknesses are either used to assemble detuned control cavities or are stripped with toluene so the DBRs can be reused. We additionally prepare extracavity thin films on 1" diameter, 3 mm thick UV-FS substrates for control experiments using an identical procedure.

*2.3.4. Cavity assembly.* To construct two-mirror FP optical cavities filled with Ce6T/PS, we use a NX-2000 nanoimprinter to thermally bond a bare ½" DBR on top of a spin-coated 1" DBR. Figure 2c shows experimental and simulated TMM transmission spectra for a representative "empty" cavity containing a 711 nm PS film, which supports a single longitudinal cavity mode in the region of the DBR stop band. We report an experimental empty-cavity mode full-width at half-maximum (fwhm) linewidth of ~50 meV, a close match for the 66 meV fwhm absorption linewidth of the Ce6T $S_0 \rightarrow Q_y$ transition to be coupled.

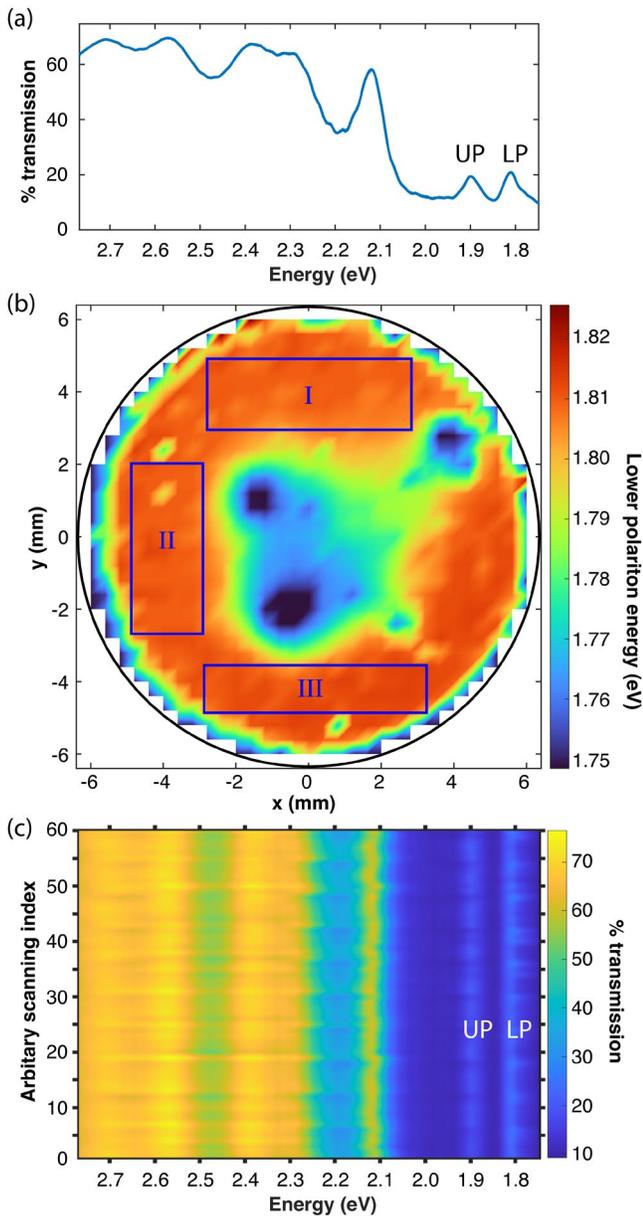

**Figure 3.** Characterization of spatial uniformity in Cavity 1, which is filled with an 809 nm thick Ce6T/PS film that reaches the strong coupling regime in the $Q_y$ band. (a) Cavity transmission spectrum of Cavity 1 at normal incidence in the strongly coupled region. Upper and lower polariton (UP, LP) features are visible at 1.81 and 1.90 eV, respectively. (b) Spatial cavity-coupling map tracking the energy of the lower polariton feature near 1.81 eV as a function of $x$ and $y$ coordinates in the cavity plane. Regions of uniform cavity-coupling are highlighted in blue boxes labeled I, II, and III. (c) Cavity transmission spectra acquired while rastering over region I of the map shown in panel (b). The cavity-coupling conditions do not change significantly as a function of spatial coordinates, indicating that rastering can be performed over this region without significant cavity detuning.

*2.3.5. Ce6T solutions.* We prepare solutions of 10 μM and 100 μM Ce6T in toluene to make solution-phase reference measurements for comparison with thin film results. We use the lower concentration solution in a 2 mm pathlength quartz cuvette (Ossila IR Quartz Cuvette) for UV-Vis and PL measurements. For pump-probe measurements, we use the higher concentration sample in a demountable flow cell (DLC-M25, Harrick Scientific) with a 500 μm sample pathlength. We flow the solution during pump-probe experiments using a peristaltic pump (Masterflex) and Viton tubing with an inner diameter of 0.80 mm (Avantor).

**2.4. Accounting for cavity nonuniformity in rastering.** In performing ultrafast spectroscopy on Ce6T films, we observe that differential pump-probe signals decay in amplitude due to photodegradation over the course of a few minutes of lab time. It is therefore essential to refresh the sample by rastering, though this introduces some additional challenges for intracavity measurements. In order to maintain spatial overlap of the pump and probe beams in the sample throughout the raster, the sample must be translated in its own $x$-$y$ plane orthogonal to the probe beam. As a result, in order to perform pump-probe measurements in resonant strongly-coupled cavities, we must fabricate cavities that demonstrate strong coupling at normal incidence. This constraint underscores the importance of controlled spin coating of thin films as described above in Section 2.3.

Our DBR FP cavities exhibit spatial variation in coupling conditions across the $x$-$y$ cavity plane due to slight nonplanarity of the thin films and imperfections in the mirror bonding process. To control for this, we create a spatial cavity-coupling map for each device to identify uniform regions amenable to rastering. We acquire broadband transmission spectra with the WLC probe while rastering across the face of each cavity, and tag each spectrum with $(x, y)$ coordinates. We then plot the energy of a particular spectral feature – either a polariton or an uncoupled cavity mode – as a function of these $(x, y)$ coordinates to yield a 2D map. Figure 3a shows a linear transmission spectrum of Cavity 1 where the $S_0 \rightarrow Q_y$ transition is coupled to the fourth-order longitudinal cavity at normal incidence producing LP and UP features at 1.81 and 1.90 eV. By plotting the energy of the LP as a function of $(x, y)$ coordinates, we obtain the cavity coupling map in Fig. 3b. Similar maps are provided for Cavities 2 and 3 in Section S3 of the SI. Regions of the map in Fig. 3b that demonstrate relatively uniform coupling conditions are highlighted in blue boxes. Fig. 3c plots the transmission spectra acquired while rastering across region I of Cavity 1. All features in these transmission spectra – including the UP, LP, and an uncoupled cavity mode near 2.1 eV – exhibit minimal dispersion throughout the raster. All intracavity pump-probe data reported herein are collected while rastering over regions identified as being similarly spatially uniform.

## 3. RESULTS

We now turn to experimental characterization of Ce6T photophysics in free space and under ESC. First, we characterize the linear absorption and PL spectra of extracavity Ce6T in Section 3.1. We then present pump-probe and TCSPC data for extracavity Ce6T in Section 3.2. In Section 3.3 we demonstrate strong coupling of Ce6T in both Cavities 1 and 2. Finally, we present direct pump-probe measurements of the excited-state dynamics of Ce6T under ESC in Section 3.4.

**3.1. Absorption and PL of extracavity Ce6T.** We begin by reviewing the photophysics of Ce6T in free space. Here, we present linear absorption and PL spectra for a dilute 10 μM solution of Ce6T in toluene and a concentrated Ce6T/PS thin film.

The absorption spectrum of dilute solution-phase Ce6T is shown in blue in Fig. 4a. Three absorption bands are evident centered at 1.86 eV, 2.46 eV, and 3.07 eV. These bands can be described with the four-orbital model commonly used to interpret spectra of porphyrins.[102] The strong band at 3.07 eV is the sum of two near-degenerate transitions collectively labeled as the B band. The weaker features at 1.86 eV and 2.46 eV are referred to as the $Q_y$ and $Q_x$ bands, respectively. These transitions are formally forbidden, but draw oscillator strength from the B band via vibronic coupling.[102]

Figure 4b plots the PL spectrum of solution-phase Ce6T after excitation of the B band. The spectrum is dominated by emission from $Q_y$ at 1.84 eV. It is known that chlorins exhibit rapid (<1 ps) internal relaxation of the B band, yielding fluorescence primarily from the lowest-lying $Q_y$ band.[88,90,93,95,103] Previous literature reports solution-phase $Q_y$ lifetimes that persist for several nanoseconds, with radiative relaxation and intersystem crossing to the ground state the dominant relaxation channels.[18,89,90,92,95,104]

We next characterize concentrated thin film samples of Ce6T in PS. The absorption spectrum of Ce6T/PS films (red trace in Fig. 4a) is nearly identical to that seen in solution apart from the appearance of a new weak band at 1.69 eV. This peak may be a sign of minor aggregation of ground-state Ce6T molecules in the concentrated film.[103,105–107] More striking deviations from the dilute solution-phase results manifest in the Ce6T thin film PL spectra after excitation of the B band at 3.10 eV (red trace in Fig. 4b). A weak band is evident at 1.82 eV which we attribute to partly obscured emission from the same $Q_y$ band that characterizes the solution-phase PL spectrum. The thin film spectrum is dominated, however, by a feature at 1.66 eV, which we assign to Ce6T excimer emission, in line with previous reports.[18,19] Excimers are delocalized excitations shared between an excited state molecule and nearby ground-state molecules that interact strongly in a collective excited state.[108–110] In dense thin films, Ce6T molecules therefore appear to interact and aggregate only weakly in the ground state but interact strongly to form excimers upon electronic excitation.

### 3.2. Ultrafast pump-probe spectroscopy and TCSPC of extracavity Ce6T.
We now discuss the excited state dynamics of extracavity Ce6T. Pump-probe and TCSPC data for solution-phase Ce6T in toluene are presented in Section S4 of the SI. We find that the $Q_y$ state of Ce6T is long-lived in dilute solution, consistent with the literature.[18,19] We provide these solution-phase results as a reference point, but focus the remainder of this section chiefly on the Ce6T/PS thin film data to provide a more relevant benchmark for studies of intracavity films. In Section 3.2.1, we use pump-probe measurements to track the $Q_y$ state and excimer populations in thin films via the broad Ce6T ESA feature. In Section 3.2.2, we obtain isolated signatures of the excimer dynamics by probing the excimer emission with both TCSPC and pump-probe experiments.

*3.2.1. $Q_y$ excited state dynamics in Ce6T/PS films.* Figures 5ab show pump-probe spectra for a Ce6T/PS thin film following optical pumping of the B band at 3.10 eV (400 nm) with the static absorption spectrum plotted in Fig. 5c for reference (reproduced from Fig. 4a). This excitation scheme indirectly populates the $Q_y$ manifold via rapid, sub-picosecond internal conversion.[88,89,93,94] The pump-probe spectra show a negative-going feature at 1.86 eV which corresponds to bleaching of the $S_0 \rightarrow Q_y$ band as the ground state is depleted by the pump. A broad positive-going feature spanning 2.0–2.8 eV also appears, which we assign to the excited state absorption of both the $Q_y$ excited state and excimers (*vide infra*).

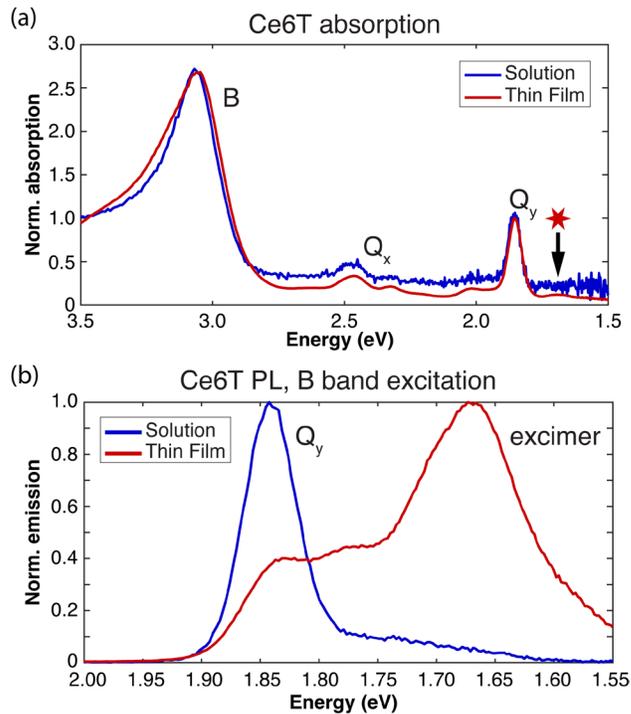

**Figure 4.** Optical absorption and photoluminescence (PL) spectra of Ce6T in 10 μM toluene solution (blue traces) and 16 wt. % thin PS films (red traces). (a) The linear absorption spectra for Ce6T in solution and thin film are nearly identical, apart from a weak absorption feature at 1.69 eV that appears only in the thin film (red star). (b) PL spectra for Ce6T in solution and thin film acquired with excitation at 3.06 eV. In solution, emission primarily occurs from the $Q_y$ band while red-shifted emission from the excimer dominates in thin films.

Figure 5d shows a temporal linecut of the $Q_y$ ESA signature averaged over the spectral window from 2.35–2.42 eV. Trimming the data before 1 ps, the subsequent dynamics are well-fit to three parallel exponential decays. We compare fits for three spectral windows across the broad ESA (2.20–2.32 eV, 2.35–2.42 eV, and 2.55–2.75 eV) and find that all windows feature the same dynamics, consistent with previous reports.[19,95] We report three ESA decay time constants of $\tau_1 = 7.0 \pm 1.4$ ps, $\tau_2 = 57 \pm 6$ ps, and $\tau_3 = 1110 \pm 100$ ps (Table 1), which represent the average of fits performed for all three spectral windows across 26 data sets collected for six different thin film samples pumped at 3.10 eV. We also fit the recovery of the $S_0 \rightarrow Q_y$ bleach with a three-exponential model, finding time constants of $6.5 \pm 0.8$ ps, $49 \pm 6$ ps, and $1050 \pm 150$ ps. The ESA decay and $Q_y$ bleach recovery timescales are consistent within experimental uncertainty, suggesting that long-lived triplet or other intermediate states are insignificant in Ce6T thin films.

We observe the same ESA dynamics when we pump the $Q_y$ band directly at 1.85 eV (669 nm), and when we excite to the red or the blue of the $Q_y$ band at 1.81 eV (684 nm) or 1.90 eV (653 nm), respectively. These latter two detuned pump energies represent extracavity controls for optical excitation of the lower and upper polariton in intracavity samples, as we discuss later in Section 3.4. These results are summarized in Table 1 and additional representative spectra are provided in Section S5 of the SI.

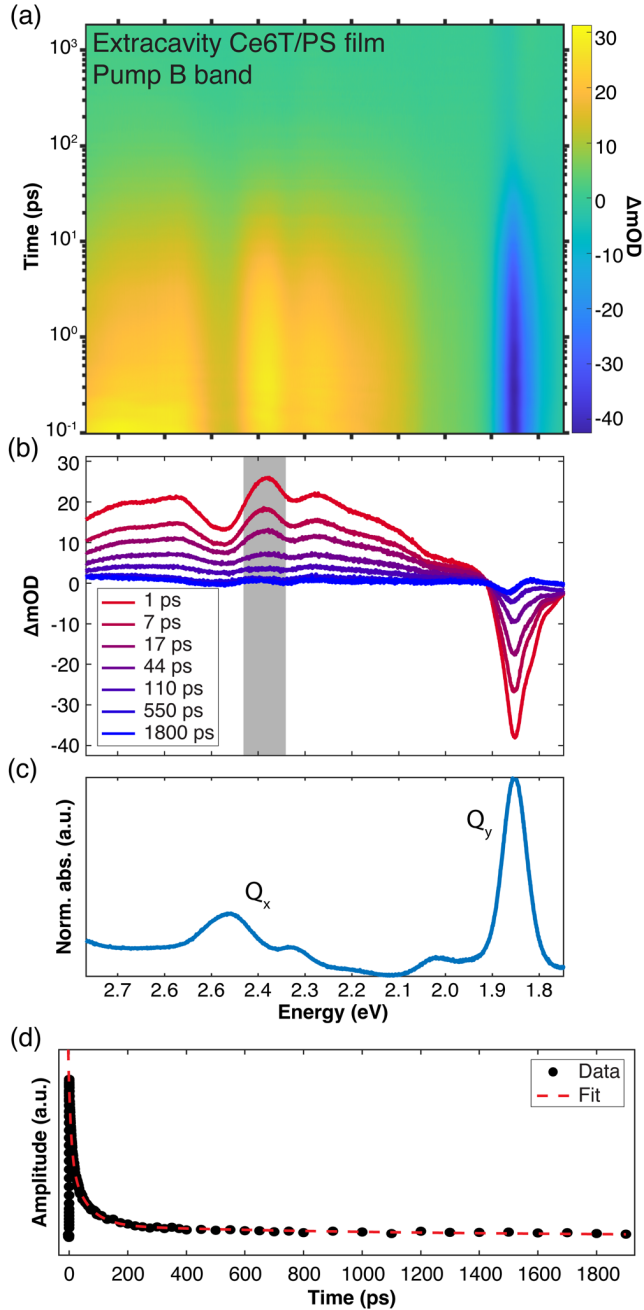

**Figure 5.** Transient dynamics of an extracavity Ce6T/PS film following optical excitation of the B band at 3.10 eV (400 nm). (a) Broadband pump-probe spectra and (b) representative spectral linecuts. (c) Linear absorption spectrum of Ce6T/PS replotted from Fig. 4a to illustrate where relevant spectral features lie. (d) Temporal linecut of the pump-probe data from panel (a) showing the ESA dynamics averaged over the spectral window from 2.35–2.42 eV (as marked in gray in panel (b)). Experimental data points are shown with black dots, while the red dashed line represents a fit of these data to three parallel exponential decays.

The observed $\tau_1$, $\tau_2$, and $\tau_3$ ESA decay constants likely stem from a range of processes, including vibrational cooling, excimer formation, and internal conversion of both excited-state monomers and excimers back to the ground state. We also suspect that morphological inhomogeneities in the Ce6T/PS films play a role, and that, as others have noted,[28,111] molecules residing in different microenvironments within thin films exhibit a distribution of time constants. The shortest $\tau_1$ decay constant is commensurate with the reported vibrational cooling times for porphyrinoid chromophores,[88,89,93,103,105,106] but is also close to the excimer formation time (*vide infra*) and may therefore arise from a combination of both processes. The longer $\tau_2$ and $\tau_3$ constants are in the range expected for internal conversion lifetimes for aggregated porphyrinoids[89,103,105,106] and are also consistent with excimer relaxation lifetimes (*vide infra*). Again, our fitted time constants likely arise from a combination of these processes.

*3.2.2. Excimer dynamics in Ce6T/PS films.* We perform additional extracavity experiments to understand the excimer dynamics in Ce6T/PS films, as these delocalized states dominate the static PL emission in concentrated samples and likely contribute to the ESA dynamics discussed above. We first perform TCSPC measurements in Ce6T/PS thin films, exciting the B band at 3.09 eV (401 nm) and probing the excimer emission at 1.68 eV (740 nm). The results are laid out in Section S6 of the SI. We find that excimer formation occurs well within the IRF of the TCSPC instrument while excimer relaxation is well described by two parallel exponential decays with time constants of 230 ± 30 ps and 1170 ± 120 ps.

We use ultrafast pump-probe spectroscopy to corroborate the TCSPC excimer measurements, as detailed further in Section S6 of the SI. In these experiments, we optically excite the Ce6T/PS film at 3.10 eV and examine the excimer stimulated emission (SE) which appears as a negative-going feature at 1.67–1.72 eV (the same spectral location as the excimer PL emission). We find that a temporal linecut of this SE feature is well-fit by a sum of one rising exponential with time constant 7.9 ± 0.5 ps and two decaying exponentials with time constants 240 ± 30 ps and 1300 ± 110 ps. These results are in excellent agreement with the TCSPC data. We do note that our analysis of the excimer kinetics differs somewhat from the recent work of Biswas *et al.*[18] We provide more context and discussion of this topic in Section S6 of the SI.

In any event, the observed excimer kinetics are straightforward to interpret. The 7.9 ps rise time is consistent with literature lifetimes for the formation of non-diffusion-limited excimers.[109,112–114] The two decay time constants likely stem from morphological inhomogeneities in the sample which give way to varying degrees of excimer aggregation and thus varying non-radiative relaxation lifetimes.[28,114] We provide a diagrammatic summary of the dynamics we observe in Section S7 of the SI.

**3.3. Strong coupling the Ce6T $Q_y$ band in DBR cavities.** We now demonstrate strong coupling of the $S_0 \rightarrow Q_y$ transition of Ce6T in two DBR cavity devices. Both cavities are constructed from the dichroic DBRs detailed in Section 2.2 and feature reflective stop bands centered at 1.86 eV to allow for strong coupling of the Ce6T $Q_y$ band. Cavity 1 contains an 809 nm thick Ce6T/PS film, such that the fourth-order longitudinal cavity mode is resonant with the Ce6T $Q_y$ band near normal incidence, while Cavity 2 contains a 979 nm thick Ce6T/PS film, with the fifth-order longitudinal cavity mode resonant with the Ce6T $Q_y$ band at normal incidence.

An experimental transmission spectrum of Cavity 1 is shown in blue in Fig. 6a; LP and UP bands are evident at 1.81

and 1.90 eV. The Rabi splitting for this device is 86 ± 4 meV at normal incidence, which exceeds both the Ce6T $Q_y$ band linewidth of 66 meV fwhm and the empty cavity linewidth of 50 meV fwhm. The stated uncertainty in the Rabi splitting represents the standard deviation over splittings recorded as we raster across the cavity; see the spatial cavity-coupling map of this device in Fig. 3. The dashed red line in Fig. 6a plots a TMM simulation of this device. The background refractive index, $n_0$, and Ce6T concentration, [Ce6T], are fit to the experimental data using the $Q_y$ band extinction coefficient from the literature (34,911 M$^{-1}$cm$^{-1}$),[18] yielding $n_0 = 1.65$ and [Ce6T] = 0.16 M. A dispersion plot of the experimental transmission of Cavity 1 as a function of angle is shown in Fig. 6b, evidencing the characteristic avoided crossing behavior of the polariton bands as they pass through resonance with the $Q_y$ band at 1.86 eV.

Cavity transmission and dispersion spectra for Cavity 2 are presented in Figs. 6cd; the spatial cavity-coupling map for this device is also provided in Section S3 of the SI. Cavity 2 features a Rabi splitting of 128 ± 2 meV, significantly larger than that achieved in Cavity 1. We ascribe this to higher intracavity [Ce6T] in Cavity 2, as this device was spin coated with Ce6T/PS/toluene solution drawn from the bottom of a batch, where material had settled. We perform control experiments in Cavity 2 as a means of testing whether the collective coupling strength and/or coupled cavity mode order have any impact on intracavity dynamics.

**3.4. Pump-probe measurements of Ce6T in DBR cavities.** We now detail the results of ultrafast pump-probe experiments performed for intracavity Ce6T/PS films under resonant strong coupling of the $Q_y$ band in DBR cavities. The chief aim of these experiments is to record the excited state lifetimes of the strongly-coupled $Q_y$ manifold following both indirect population of $Q_y$ via pumping the higher-lying B band and direct optical pumping of the $Q_y$ band itself. By design, the high transmission of the DBR mirrors from 2.2–2.8 eV permits spectroscopic access to ESA signatures in pump-probe measurements.

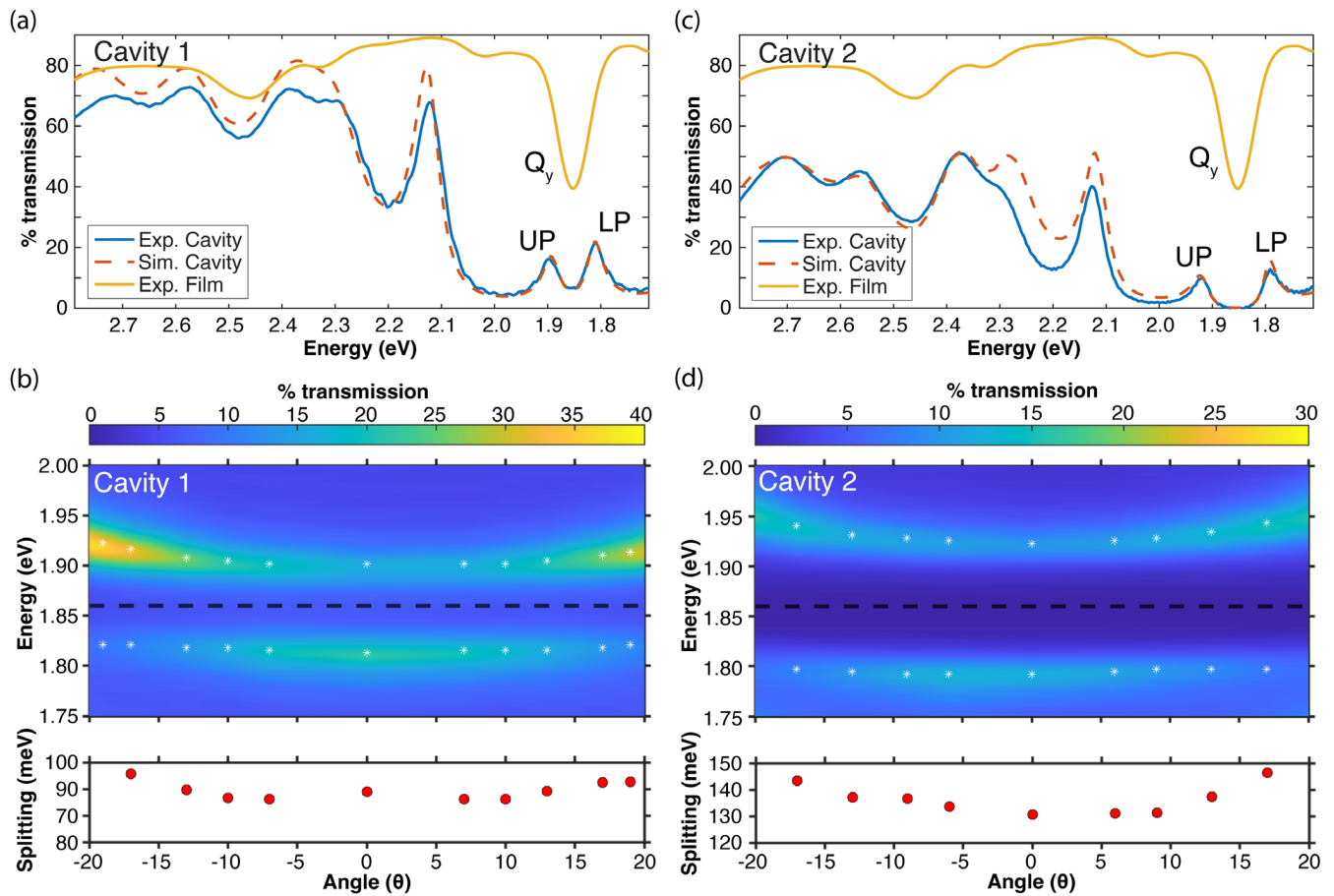

**Figure 6.** Strong coupling of the $Q_y$ band of Ce6T in PS films embedded in Cavities 1 and 2 (809 nm and 979 nm thick). (a,c) Experimental normal incidence transmission spectra (blue traces) of Cavities 1 and 2 show formation of upper and lower polaritons (UP, LP) centered about the Ce6T $Q_y$ band. Simulated TMM cavity transmission spectra (dashed orange traces) are fit to experimental data by floating the background refractive index and Ce6T concentration giving $n_0 = 1.65$ and [Ce6T] = 0.16 M for Cavity 1 and $n_0 = 1.67$ and [Ce6T] = 0.51 M for Cavity 2. Experimental transmission spectra through extracavity Ce6T/PS films are plotted in yellow for reference. (b,d) Angle-tuned dispersion spectra of Cavities 1 and 2 show the characteristic avoided crossing behavior of the UP and LP features at normal incidence in both structures, where the Rabi splitting is minimized, as plotted with red dots.

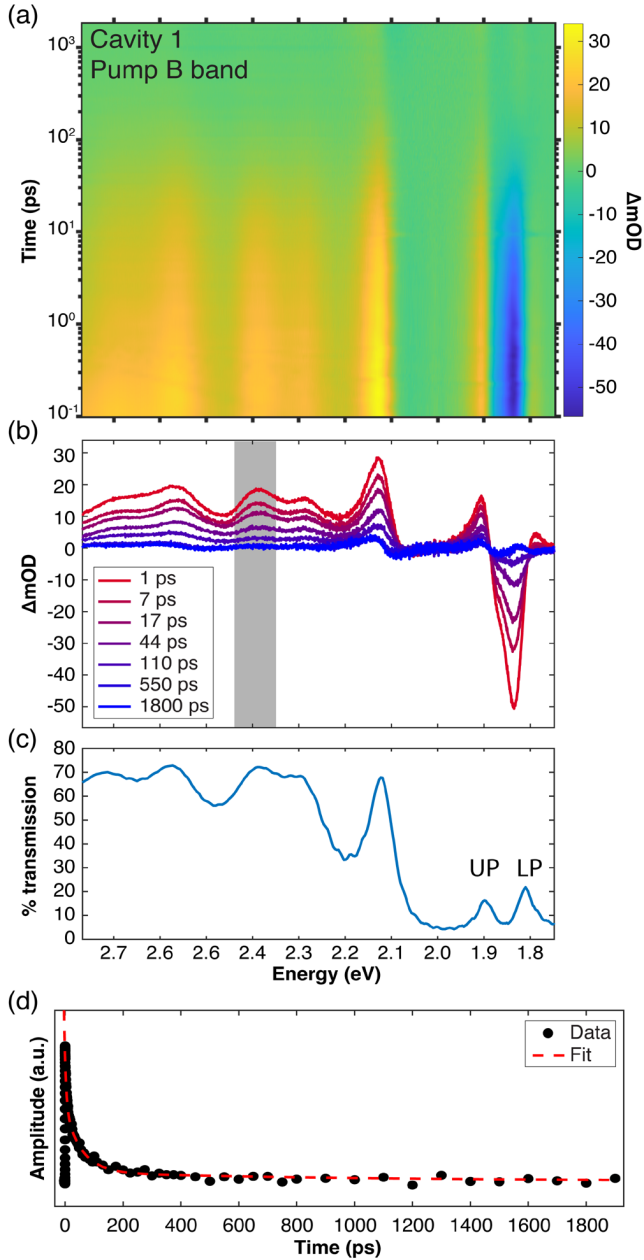

transient features. Similar measurements are provided for Ce6T/PS films embedded in Cavity 2 and detuned Cavity 3 in Section S8 of the SI.

Optical pumping of the intracavity systems at 3.10 eV proceeds much the same as it does for extracavity samples given the high transmission of the DBR cavities in this spectral region (Fig. 2). We therefore expect that excited population in the B band should rapidly relax into the cavity-coupled $Q_y$ band, indirectly populating the polariton manifold.[8,115–117] Indeed, the pump-probe spectra at positive delay times are dominated by the ESA feature, which appears clearly from 2.2–2.8 eV in Figs. 7ab through the high transmission window of the DBRs. We fit the ESA decay lifetimes for both Cavities 1 and 2 using the same methodology employed in extracavity experiments (see Section 3.2): temporal linecuts are constructed by averaging over three separate spectral windows across the ESA (2.20–2.32 eV, 2.35–2.42 eV, and 2.55–2.75 eV). These linecuts are then fit to a parallel three-exponential decay to extract time constants. All fitted time constants found in resonant strongly-coupled cavities pumped at 3.10 eV are reported in Table I, alongside control results obtained in extracavity films and detuned cavities. Within experimental error, we find identical excited state dynamics in all cases. Cavity coupling therefore does not appear to affect any observed ESA decay constants following indirect excitation of the polariton manifold via pumping of the Ce6T B band.

While the ESA region of the transient spectrum of strongly-coupled Ce6T in Figs. 7ab looks much like that of the extracavity system, it is worth commenting on the distinct transient signatures in the DBR stop band. At positive delay times, two derivative-like features appear on either side of the $Q_y$ band at 1.86 eV in Figs. 7ab. These features arise due to the well-known transient contraction of the collective Rabi splitting which occurs as bleaching of the $S_0 \rightarrow Q_y$ transition reduces the number of intracavity molecules available for cavity coupling. The Rabi splitting recovers as the system relaxes and the ground state is repopulated. A similar sharp derivative-like feature is also evident near 2.1 eV, which decays on a timescale commensurate with the ESA. This feature coincides with the position of an uncoupled cavity mode, which appears to shift transiently as the intracavity background refractive index is modulated by optical pumping. Each of these spectral features appear to decay with the same timescales as the ESA and $Q_y$ band bleach. The persistence of these transient cavity features in the DBR stop band illustrates the difficulties inherent in performing pump-probe measurements through reflective optics[76] and the comparative clarity of the dichroic ESA readout.

*3.4.2. Pumping the strongly-coupled $Q_y$ band.* We next turn to examine excited state lifetimes following direct optical excitation of the strongly-coupled $Q_y$ band. We perform three sets of pump-probe experiments in Cavity 1, tuning the pump excitation energy to excite the $Q_y$ band at 1.85 eV as well as to directly excite the lower and upper polaritons at 1.81 eV and 1.90 eV.

Pumping at 1.85 eV should directly excite the intracavity molecular reservoir at the bare excitation frequency, relying on partial transmission of pump light through the input cavity mirror.[47] Meanwhile, excitation at 1.81 and 1.90 eV should coherently excite the polaritonic modes, which are then expected to dephase within the cavity lifetime of ~13 fs, as calculated from the empty cavity mode linewidth. Representative

**Figure 7.** Transient dynamics of a Ce6T/PS film under strong coupling of the $Q_y$ band in DBR Cavity 1 following optical excitation of the B band at 3.10 eV (400 nm). (a) Broadband pump-probe spectra and (b) representative spectral linecuts. Excited state absorption (ESA) features are clearly visible from 2.2–2.8 eV through the transparent region of the DBR mirrors. Rabi contraction of the polariton bands is also evident near 1.86 eV. (c) Linear transmission spectrum of Cavity 1 replotted from Fig. 6a. (d) Temporal linecut of the pump-probe data from panel (a) showing the ESA decay averaged over the spectral window from 2.35–2.42 eV (as marked in gray in panel (b)). Experimental data points are shown with black dots, while the red dashed line represents a fit of these data to three parallel exponential decays.

*3.4.1. Pumping the Ce6T B band.* Figures 7ab show the transient white-light-probe transmission spectrum of a Ce6T/PS film strongly-coupled in Cavity 1 following optical pumping of the B band at 3.10 eV. These spectra are plotted against the linear transmission spectrum of Cavity 1 in Fig. 7c (reproduced from Fig. 6a) for ease of comparison with transient data are provided in Fig. 8 for excitation of Cavity 1 at 1.85 eV, and in Section S9 of the SI for direct excitation of the polariton bands.

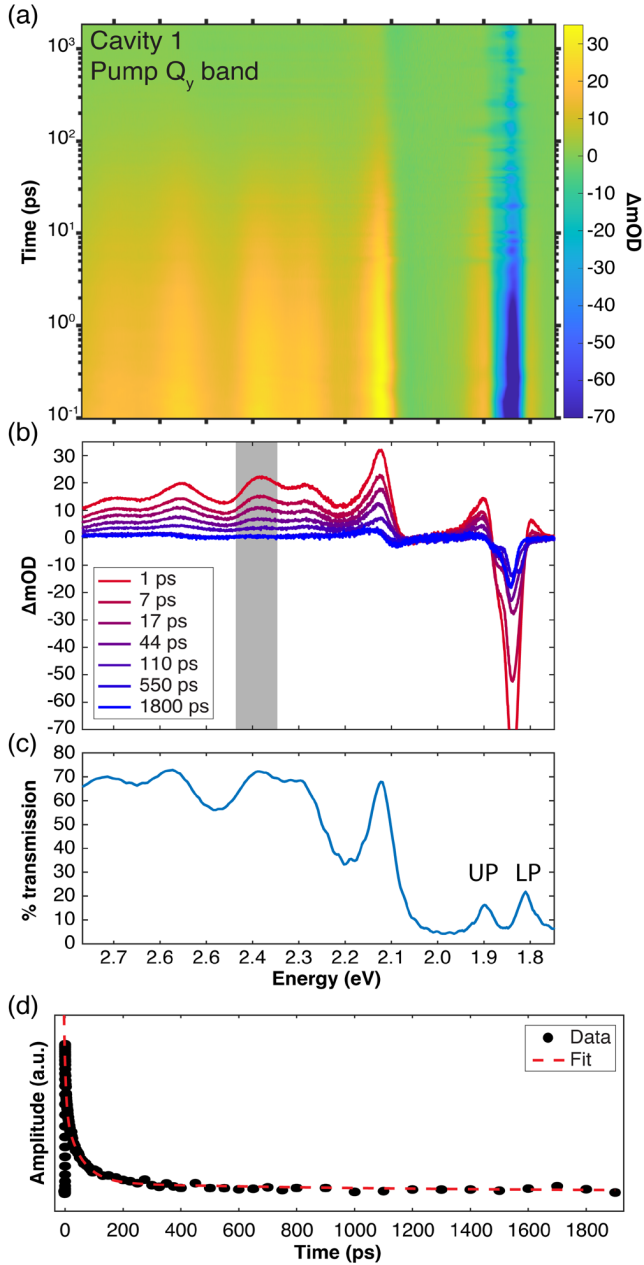

**Figure 8.** Transient dynamics of a Ce6T/PS film under strong coupling of the $Q_y$ band in DBR Cavity 1 following optical excitation of the $Q_y$ band at 1.85 eV (669 nm). (a) Broadband pump-probe spectra and (b) representative spectral linecuts. Excited state absorption (ESC) features are clearly visible from 2.2–2.8 eV through the transparent region of the DBR mirrors. Rabi contraction of the polariton bands is also evident near 1.86 eV. Noise in the transient optical density at 1.85 eV derives from scattered pump light. (c) Linear transmission spectrum of Cavity 1 replotted from Fig. 6a. (d) Temporal linecut of the pump-probe data from panel (a) showing the ESA decay averaged over the spectral window from 2.35–2.42 eV (as marked in gray in panel (b)). Experimental data points are shown with black dots, while the red dashed line represents a fit of these data to three parallel exponential decays.

**Table I.** Time constants for excited state lifetimes in extracavity Ce6T/PS films, Ce6T/PS films in detuned FP cavities, and Ce6T/PS films under resonant strong coupling of the $S_0 \rightarrow Q_y$ transition, following optical excitation with various pump energies [a]

| Pump: 3.10 eV (400 nm) | $\tau_1$ (ps) | $\tau_2$ (ps) | $\tau_3$ (ps) |
|---|---|---|---|
| extracavity | 7.0 ± 1.4 | 57 ± 6 | 1110 ± 100 |
| off-resonance | 6.6 ± 1.6 | 69 ± 12 | 1200 ± 400 |
| resonant polariton | 6.4 ± 1.7 | 58 ± 8 | 1100 ± 200 |
| Pump: 1.85 eV (669 nm) | $\tau_1$ (ps) | $\tau_2$ (ps) | $\tau_3$ (ps) |
| extracavity | 6.7 ± 1.3 | 59 ± 6 | 1060 ± 50 |
| resonant polariton | 6.3 ± 2.0 | 64 ± 15 | 1070 ± 120 |
| Pump: 1.81 eV (684 nm) | $\tau_1$ (ps) | $\tau_2$ (ps) | $\tau_3$ (ps) |
| extracavity | 5.9 ± 1.9 | 62 ± 9 | 1070 ± 70 |
| resonant polariton | 6.4 ± 1.8 | 62 ± 7 | 1040 ± 30 |
| Pump: 1.90 eV (653 nm) | $\tau_1$ (ps) | $\tau_2$ (ps) | $\tau_3$ (ps) |
| extracavity | 5.5 ± 1.5 | 61 ± 10 | 1070 ± 100 |
| resonant polariton | 6.7 ± 2.0 | 65 ± 10 | 1020 ± 30 |

[a] More information regarding how many data sets were averaged for each condition, how many film samples were tested for each condition, results for individual data sets, and the goodness of each fit can be found in the SI spreadsheet. The error bars on each time constant represent one standard deviation of the fit results across all data fit for that condition.

All pump-probe spectra acquired with optical excitation in the $Q_y$ band region are qualitatively similar to those acquired pumping the B band. The transient spectra plotted in Fig. 8 and SI Section S9 feature the same broad ESA signature spanning 2.2−2.8 eV and a transient Rabi splitting contraction centered at the $Q_y$ band center at 1.86 eV. We again apply a parallel three-exponential model to fit the ESA temporal dynamics and report all time constants in Table I. Regardless of pump energy, we find no statistically significant differences in any decay time constants as compared to extracavity thin films. These results signal that direct optical population of the strongly-coupled $Q_y$ band does not detectably affect the excited state lifetime.

## 4. DISCUSSION

We report here on a new platform to directly probe excited state lifetimes in molecules under ESC, considering both indirect and direct optical population of the polaritonic manifold. We minimize optical filtering effects by working in purpose-fabricated dichroic DBR cavities. Examining the lifetimes of the $Q_y$ band of Ce6T under ESC as a testbed, we do not observe any significant deviations in excited state lifetimes under any cavity conditions studied herein. We now discuss these results in more detail and consider which experimental parameters may be key to observing cavity-modification of excited state dynamics. In Section 4.1, we lay out the technical considerations that facilitate the reliable quantification of Ce6T excited state lifetimes in our measurements. In Section 4.2 we consider potential explanations for negligible cavity-modification of the excited state lifetimes of Ce6T and compare our results with the literature.

## 4.1. Technical considerations for probing ultrafast dynamics under ESC.
We first discuss the challenges one encounters in making clean measurements of ultrafast dynamics in FP nanocavities designed for ESC.

*4.1.1. Sample photodegradation.* Photodegradation is a serious concern in ultrafast studies of molecular chromophores in thin films. We begin to notice depletion of both the ESA and bleach signals in extracavity Ce6T/PS thin films within 30 seconds of exposure to pump light at 3.10 eV with 150 μJ/pulse energies at 1 kHz. We characterize the timescale of photodegradation in these samples by allowing focused pump light (~200 μm diameter) to sit at a single spot and find an exponential decay of transient signal amplitudes with a time constant of ~60 seconds. Once photodegradation has occurred, the signal does not recover within hours. While the transient signatures of Ce6T become weaker with photodegradation, no new spectral features appear, indicating that whatever photoproducts result do not have spurious transients that would interfere with our readout. We therefore simply raster the sample much faster than the photodegradation rate, using no more than a 1 second dwell time at each sample spot. Although rastering works well for this application, it does introduce additional challenges for intracavity experiments in terms of maintaining consistent strong coupling conditions across a raster (*vide infra*).

*4.1.2. Maintaining cavity coupling conditions while rastering.* Photodegradation of Ce6T necessitates that we raster throughout all pump-probe measurements. However, cavity-coupling conditions may change as we raster due to non-planarity in the thin films and/or mirror substrates and imperfections in the mirror bonding process. Cavity-coupling is extremely important to control for, as several demonstrations of modified molecular behavior under strong coupling report a sensitive dependence on cavity detuning.[10,118–120]

We control for changes in cavity coupling conditions while rastering by creating a spatial map of the transmission spectrum of each device, as detailed in Section 2.4. We then identify flat regions suitable for rastering. Despite these efforts, cavity coupling conditions are never perfectly uniform throughout a raster. We use the standard deviation of the spectral position of a polariton or an uncoupled cavity mode while rastering during a pump-probe scan as a metric for the spatial uniformity of each device. This allows us to place error bars of ±4 meV and ±2 meV on the polariton peak positions in strongly-coupled Cavities 1 and 2, respectively, and error bars of ±5 meV on the photonic mode positions in detuned control cavities. These uncertainties are relatively small, representing less than 10% of the $Q_y$ band linewidth (66 meV fwhm) or typical cavity mode linewidth (50 meV fwhm).

We note that these coupling uncertainties are minor compared to those achieved in static UV-visible or Fourier transform infrared measurements of cavities where large beam sizes sample a range of inhomogeneous coupling conditions and yet cavity-modification of chemistry has still been reported.[23,67,120,121] We are therefore confident that our reported excited state lifetimes encompass averaging over only a small range of inhomogeneous cavity coupling conditions.

*4.1.3. Optical artifacts.* Optical artifacts can greatly complicate pump-probe spectroscopy of intracavity molecules. Many transient spectral features of strongly-coupled systems can be more simply ascribed to optically-filtered signatures of the intracavity molecular reservoir.[47,76,82–84] Much of the existing literature has examined transient signatures in the polaritonic regions, taking linecuts or fitting features like the Rabi contraction or mode shifting to examine the behavior of intracavity molecules.[14,19,118,120,122] By contrast, we directly probe molecular ESA signatures through transparent spectral regions of the DBR mirrors, permitting more direct comparison to extracavity measurements with less need for analysis and interpretation.

That said, spectral artifacts from the cavity mirrors and thin film matrix can still create confusion,[76,83] even in highly-transmissive mirror regions. An ultrafast pump pulse can induce coherent phonons, refractive index changes, and layer thickness changes which can create drastic transient changes in the transmission spectra of cavity devices.[12,13,35,76,83] The ultrafast signatures of these pump-induced effects can obscure the desired molecular dynamics.

We therefore perform control experiments to verify that our measurements of intracavity excited state lifetimes are not impacted by such artifacts. In particular, we report pump-probe spectra of the bare UV-FS substrate, a single DBR mirror on UV-FS, a PS thin film on UV-FS, and a DBR cavity containing only a PS thin film in Section S10 of the SI. In all control experiments, we use a 3.10 eV pump and broadband white light probe and otherwise identical conditions to the Ce6T/PS experiments described in Sections 3.2 and 3.4. Apart from small (< 3 mOD) transient artifacts near pump-probe overlap with sub-picosecond lifetimes, these controls feature no noticeable transients that could confound our experimental results. We therefore confirm that our dichroic devices are relatively immune to cavity artifacts and should provide a reliable readout of Ce6T excited state lifetimes under ESC.

## 4.2. Considerations for unaltered excited state lifetimes in cavity-coupled Ce6T.
We now consider potential explanations for why we do not observe modified excited state lifetimes in Ce6T under ESC. We consider the magnitude of the collective coupling achieved in our thin film samples, the orientational and positional disorder of Ce6T molecules within the cavity, and the competition of excimer formation with polariton dynamics. Ultimately, we can explain all intracavity results herein as the dynamics of the incoherent Ce6T reservoir filtered through the cavity transmission spectrum.

*4.2.1. Limitations in collective cavity coupling strength.* Many proposed mechanisms for altered molecular behavior under ESC rest on the scale of the Rabi splitting.[14,20,53,57,123–126] In general, we expect that a larger Rabi splitting should yield a stronger modification of intracavity dynamics.[127,128] It is possible that the Rabi splittings we achieve for Ce6T of 86 ± 4 meV in Cavity 1 and 128 ± 2 meV in Cavity 2 may be insufficient to demonstrate modified photophysics. The chief obstacle to attaining larger coupling strengths is the solubility of Ce6T and PS in the toluene spin-coating solution, along with the finite strength of the Ce6T $Q_y$ band.[18] Some previous studies of molecular photochemistry and photophysics under ESC achieve significantly larger Rabi splittings (300–700 meV),[15,25] with some even entering the ultrastrong coupling regime.[14,23,26] However, others have reported modified photochemistry under ESC with Rabi splittings comparable to those achieved here in Ce6T and in the presence of significant heterogeneous disorder.[9,10,18,19,73,118]

Follow-up experiments testing a larger span of Rabi splittings may be necessary to better understand the potential dependence of cavity-altered photophysics on collective coupling strength in Ce6T and in other systems. Such experiments will be subject to added difficulties as larger Rabi splittings are typically achieved with higher intracavity concentrations. Higher-concentration Ce6T thin films will likely feature increased ground-state aggregation as well as more dominant excimer participation in the excited state, which may obscure or outcompete polaritonic behavior (see Section 4.2.3).

*4.2.2. Uncoupled populations obscuring measurements.* Uncoupled and weakly-coupled intracavity molecules contribute to experimental signals alongside strongly-coupled molecules. For an individual molecule to be strongly-coupled, it must lie near an antinode of the cavity field, with its transition dipole oriented in the plane of the mirrors. In practice, a substantial fraction of intracavity molecules are positioned and oriented such that they do not experience strong coupling. The contribution of these uncoupled populations to pump-probe signals may be significant depending on the experimental configuration.

Consider exciting the B band of a Ce6T/PS film under ESC at 3.10 eV (see Section 2.4.1 and Fig. 7). As the DBR mirrors are not highly reflective in this pump region, the pump field passes through the cavity as a traveling wave and indiscriminately excites molecules along the whole cavity longitudinal axis – including those lying at both the nodes and antinodes of the longitudinal mode coupled to the $Q_y$ band. The response of this ensemble of strongly-, weakly-, and uncoupled molecules is subsequently probed via ESA signatures from 2.0–2.8 eV, which by design lie within the transmissive region of the DBR mirrors. The experiments in Fig. 7 therefore feature both pump and probe pulses that do not experience cavity confinement and therefore do not effectively discriminate for the transient signals of strongly-coupled molecules. The contributions from uncoupled molecules may obscure any polariton-mediated changes in excited state decay constants, yielding measurements that are indistinguishable from those recorded in free space.

These considerations are precisely why we also perform experiments pumping Ce6T/PS in the polaritonic region (see Section 3.4.2 and Section S9 of the SI). When pumping at the polariton frequencies, the pump field drives a standing wave in the cavity that leads to preferential excitation of molecules at the field antinodes, which happen to be the same class of molecules that are the most strongly-coupled.[47,129] Pumping in the polaritonic region should therefore enhance excitation of the strongly-coupled molecular population and reduce the fractional contribution of uncoupled molecules to the transient signal. Nonetheless, we still observe no significant difference in recorded excited state lifetimes under these conditions, suggesting that the strongly-coupled molecules themselves behave no differently than the extracavity control.

There is a clear tradeoff here: probing intracavity samples in transmissive spectral regions yields clean, artifact-free readouts but can simultaneously introduce contributions from uncoupled molecules. It remains an ongoing challenge to develop new readout strategies that optimize for both clarity and specificity in probing only the strongly-coupled intracavity molecules. In ongoing work, we are performing a more quantitative comparison of the contributions of coupled and uncoupled molecules in various polariton spectroscopy schemes.

*4.2.3. Excimer formation in Ce6T.* The participation of excimer states in the excited state dynamics of Ce6T adds another layer of complexity. We estimate that >90% of the excited molecules in our experiments ultimately go on to form excimers, based on both the time constants observed in our experiments and the photophysical model for Ce6 proposed by Kushida *et al*.[19] The formation of excimers leads to an energetic stabilization by ~180 meV, as estimated by the red-shifted PL from Ce6T in films as compared to dilute solution (Fig. 4). This stabilization shifts the excimer state out of resonance with the cavity mode, meaning that the majority of excited molecules rapidly decouple from the cavity. With this framework our null result is straightforward to rationalize: once the excited Ce6T population forms excimers, these species are no longer cavity-coupled, and therefore decay according to their free-space lifetimes.

In general, any photoinduced processes that drive strongly-coupled molecules out of the Franck-Condon region may outcompete and thereby suppress cavity-modification of photophysics. It remains an unresolved question how polariton-modified dynamics can persist in systems where the majority of molecules rapidly detune themselves from resonance with the cavity. ESC studies in molecular systems that feature aggregation, relaxation to triplet states, and photoisomerization are all subject to this caveat.[9,11,16,18–21,23,25,81,130] Aggregation in particular would appear to be a limiting factor in observing modified photophysics, since high molecular concentrations are needed to achieve ESC. At the same time, aggregates may present new opportunities, since their spatially delocalized excitations overlap more efficiently with a cavity mode volume. Addressing how to circumvent or exploit these effects remains an important open challenge for future work.

*4.2.4. The incoherent molecular reservoir.* Transiently excited polaritonic states also decay rapidly into the incoherent molecular reservoir, potentially preempting the influence of excimer formation. Recent work has argued that one should expect the polariton states to dominate photophysics only so long as there is sustained phase coherence between the coupled molecular dipoles and cavity field. In other words, the cavity photon lifetime must be commensurate with the timescale of the molecular dynamics of interest.[78,98] The cavity photon lifetime here – as is the case for most ESC studies in the literature – is on the order of tens of femtoseconds, meaning that polariton excitations decay into the incoherent reservoir well before the dynamics of interest take place. This timescale mismatch provides a simple explanation for the absence of cavity-modification of the picosecond-scale excited state dynamics of Ce6T. At the same time, literature reports of long-lived polaritonic effects under comparable conditions remain difficult to reconcile.[9,18,19]

So long as the incoherent reservoir dominates, transient cavity spectra can be described using classical optics with no need to invoke cQED. It is already well known that linear polariton spectra can be well-captured with TMM or the equivalent FP cavity expressions.[8,23,26,47,50,96,97,131–134] Recent investigations highlight that nonlinear cavity transmission spectra acquired on timescales longer than the photonic cavity lifetime can likewise be well-captured using TMM or FP techniques performed at each time step.[84,86] In other words, transient cavity spectra can often be understood as arising from the free-space molecular response filtered through the optical cavity transmission spectrum.

Our direct readout of the excited state dynamics of Ce6T via its ESA circumvents the fitting necessary to extract the dynamics from the cavity transmission spectra, instead allowing us to directly monitor the photophysics of the incoherent reservoir. We conclude that the incoherent reservoir overwhelmingly governs our ultrafast dynamics under ESC, and that classical optics provides an adequate and general framework for the interpretation of our results.

## 5. CONCLUSION

Modifying excited state molecular dynamics via strong light-matter coupling remains a tantalizing prospect. Here, we directly examine the photophysics of chlorin e6 trimethyl ester chromophores under resonant cavity coupling of the $Q_y$ excited state. We construct FP cavities using custom-fabricated dichroic DBR mirrors to achieve direct spectroscopic readout of intracavity molecular dynamics free of optical artifacts. We observe no statistically significant differences in the excited state lifetimes of intracavity molecules as compared to free-space measurements over a range of optical excitation and cavity-coupling conditions.

We discuss several potential causes for the lack of cavity-modified behavior in this system including insufficient light-matter coupling strengths, excimer formation, and contributions of both uncoupled molecules and the incoherent reservoir to transient signals. We ultimately rationalize our experimental signals as dominated by excited polaritonic molecules relaxing into the incoherent reservoir and/or funneling into excimers states that rapidly decouple from resonance with the cavity. Moving forward, it will be important for the community to reconcile reports of cavity-modified chemistry and dynamics on long (>1 ps) timescales with the emerging understanding that dynamics are unlikely to be perturbed once molecules leave the Franck-Condon region or relax into the incoherent reservoir.

We hope that the methodology we introduce here will serve as a standard for future experiments targeting the photophysics of intracavity molecules. We urge other researchers in the field to consider cavity architectures that permit clean spectral readouts in strongly coupled systems, to perform thorough control experiments, and, as much as possible, to use classical optical cavity physics models to interpret transient spectroscopic data before invoking more exotic quantum optics formalisms. In future work, we plan to apply similar methodology to investigate ultrafast electron transfer and photoisomerization reactions under ESC, as well as faster processes that compete with the cavity photon lifetime.

## ASSOCIATED CONTENT

### Data Availability Statement

Experimental details, additional plots, and experimental data are available from the authors upon reasonable request. Fits results for all pump-probe and TCSPC experiments reported in this manuscript are provided in the attached spreadsheet.

### Supporting Information

See the Supporting Information for characterization of the ultrafast pump and probe pulses, characterization of the PL IRF, ultrafast spectra of solution-phase Ce6T, cavity coupling maps for Cavities 2 and 3, steady-state and ultrafast spectra and fitting for Cavities 2 and 3, ultrafast spectra exciting at the polariton energies in Cavity 1 and extracavity films, and control pump-probe experiments with UVFS, PS films, DBR mirrors, and empty (PS-filled) cavities. Fitting results for all pump-probe and TCSPC experiments reported in this manuscript are in the attached spreadsheet.

## AUTHOR INFORMATION

### Corresponding Author

*Marissa L. Weichman – Department of Chemistry, Princeton University, Princeton, New Jersey 08544, United States; orcid.org/0000-0002-2551-9146; weichman@princeton.edu

### Authors

Alexander M. McKillop – Department of Chemistry, Princeton University, Princeton, New Jersey 08544, United States
Liying Chen – Department of Chemistry, Princeton University, Princeton, New Jersey 08544, United States
Ashley P. Fidler – Department of Chemistry, Princeton University, Princeton, New Jersey 08544, United States

### Present Addresses

†Chemistry Division, Naval Research Laboratory, Washington, DC 20375, United States

### Author Contributions

The manuscript was written through contributions of all authors. All authors have given approval to the final version of the manuscript.

### Funding Sources

This research was supported by the Air Force Office of Scientific Research (AFOSR) under grant FA9550-25-1-0157. Construction and ongoing maintenance of our ultrafast facility is supported in part with funds from the Princeton Center for Complex Materials (PCCM) NSF Materials Research Science and Engineering Center (MRSEC) under grant DMR-2011750 and by the Princeton Catalysis Initiative (PCI).

## ACKNOWLEDGMENT

We performed this work in part at the Micro/Nano Fabrication Center (MNFC), a core shared-use facility of the Princeton Materials Institute (PMI). We also acknowledge the use of the Imaging and Analysis Center (IAC) operated by PMI for our PL measurements, which is supported in part by the PCCM MRSEC. We thank Dr. Venu Vandavasi of the Princeton Biophysics Core Facility for allowing us to borrow a diode laser for TCSPC measurements. We thank Sarah Gernhart and Prof. Haw Yang for providing silica nanoparticles to measure the IRF of the PL instrument.

- SUPPORTING INFORMATION -

**Direct readout of excited state lifetimes in chlorin chromophores
under electronic strong coupling**

Alexander M. McKillop[1], Liying Chen[1], Ashley P. Fidler[1,†], and Marissa L. Weichman[1,*]

[1]Department of Chemistry, Princeton University, Princeton, New Jersey, 08544, United States

[†]Current Address: Chemistry Division, Naval Research Laboratory, Washington, DC 20375, United States

*weichman@princeton.edu

## S1. Spectral characterization of ultrafast pulses

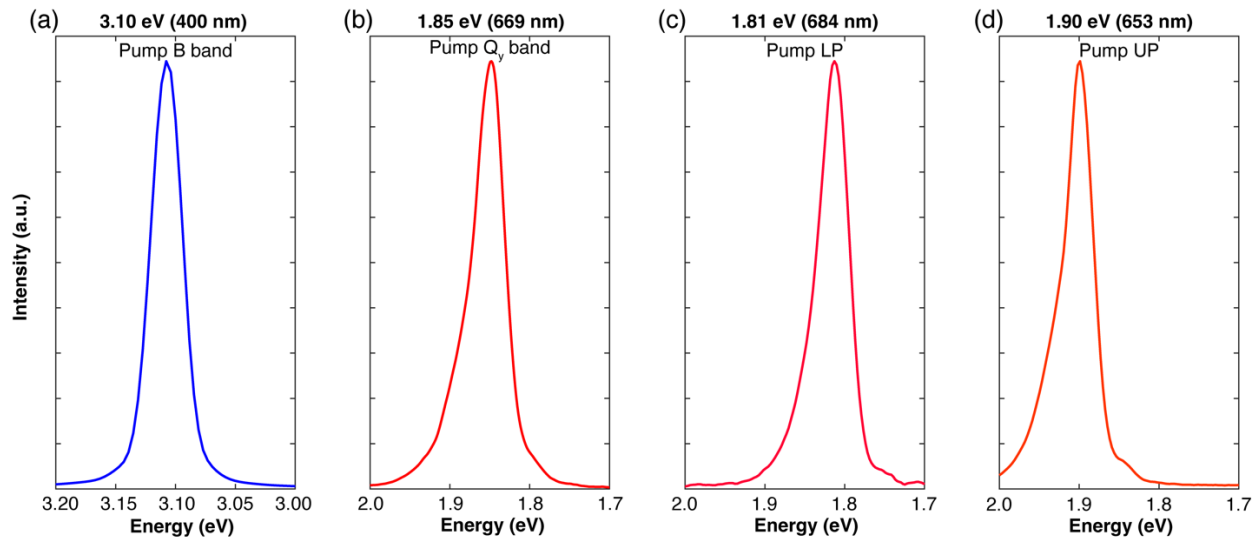

**Figure S1.** Spectra of pump pulses used in ultrafast pump-probe measurements. (a) Spectrum of 3.10 eV light used to pump the B band of Ce6T. (b) Spectrum of 1.85 eV light used to pump the $Q_y$ band of Ce6T. (c) Spectrum of 1.81 eV light used to pump the lower polariton of Ce6T in Cavity 1 and in extracavity control experiments. (d) Spectrum of 1.90 eV light used to pump the upper polariton of Ce6T in Cavity 1 and in extracavity control experiments.

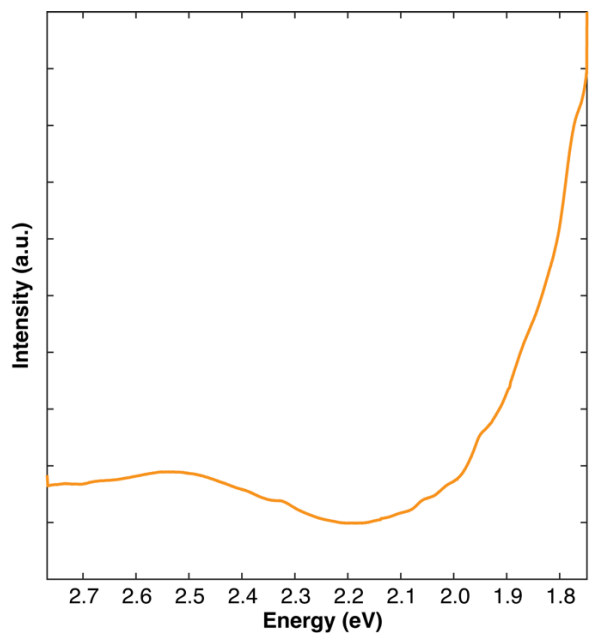

**Figure S2.** Spectrum of white light continuum probe pulses used in ultrafast pump-probe measurements.

# S2. Characterization of the TCSPC instrument response function

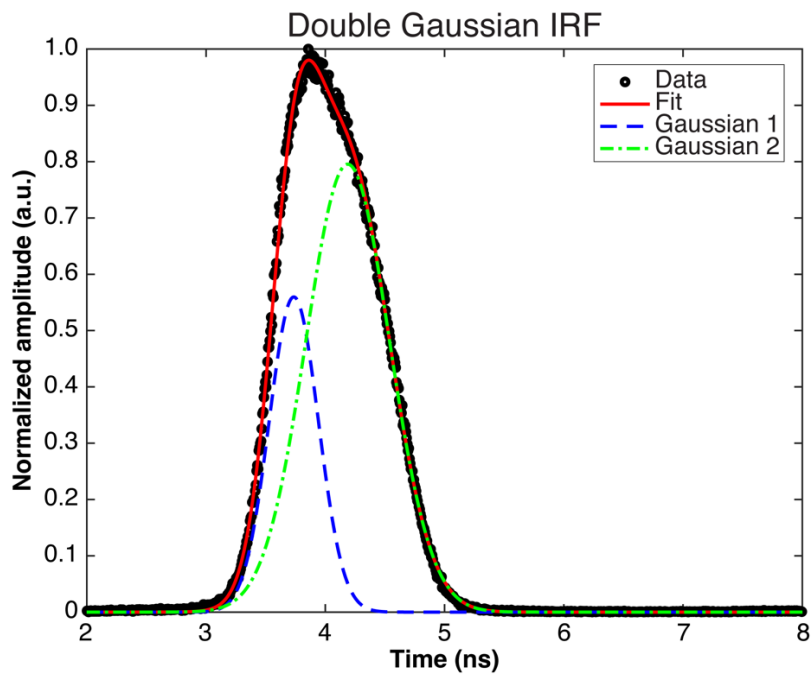

**Figure S3.** Ultrafast emission data from time-correlated single photon counting (TCSPC) scattering measurements from a 1.75mg/mL solution of 50 nm silica nanospheres in ethanol (nanoComposix). Data is collected using a photoluminescence spectrometer (FLS980, Edinburgh Instruments) fitted with a picosecond pulsed diode laser (EPL-405, Edinburgh Instruments) which excites the sample at 3.09 eV (401 nm) with a pulse width of 58.1 ps. We probe the scattered 3.09 eV light as a metric for the TCSPC instrument response function (IRF). The experimental data (black dots) is fit with a sum of two Gaussian lineshapes (red line) using the lsqcurvefit function in MATLAB. Gaussian 1 (blue dotted line) features a temporal width of 0.479 ns FWHM, while Gaussian 2 (green dotted line) is 0.810 ns FWHM in width. These two components feature relative amplitudes of 0.733:1, and Gaussian 2 is delayed from Gaussian 1 by 0.485 ns.

## S3. Spatial cavity-coupling map and transmission spectra for Cavities 2 and 3

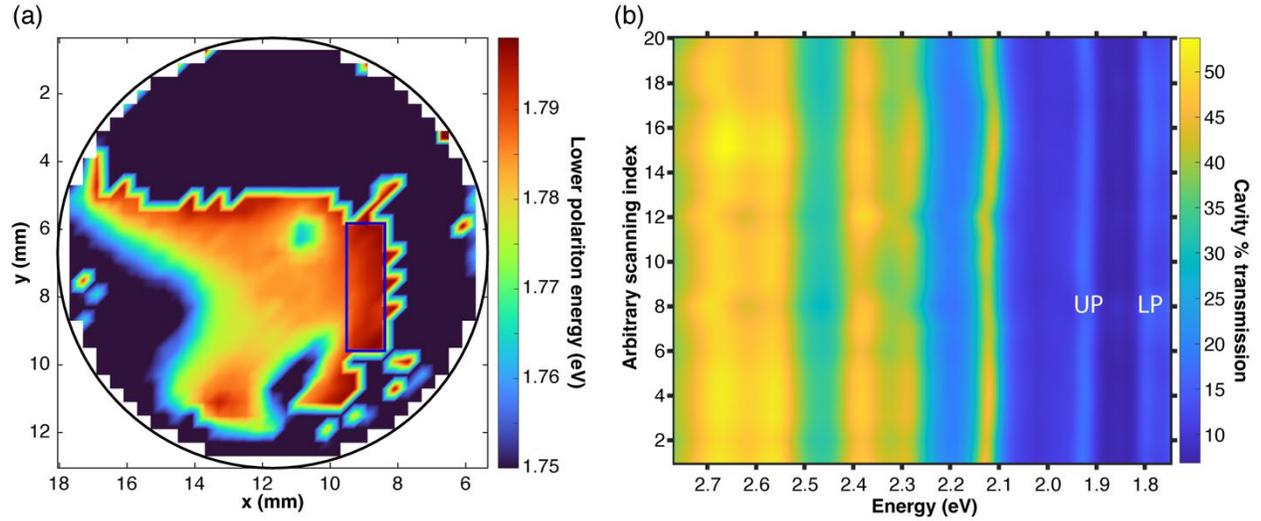

**Figure S4.** Characterization of spatial uniformity in Cavity 2, which is filled with a 979 nm thick Ce6T/PS film such that the fifth-order longitudinal cavity mode resonant strongly couples with the Ce6T $Q_y$ band near normal incidence. (a) Spatial cavity-coupling map tracking the energy of the LP as a function of $x$ and $y$ coordinates in the cavity plane. A region of uniform cavity-coupling used for pump-probe experiments is highlighted in the blue box. (b) Cavity transmission spectra acquired while rastering over the boxed region of the map shown in panel (a). The cavity-coupling conditions do not change significantly as a function of spatial coordinates, indicating that rastering can be performed over this spatial region during a pump-probe experiment without significant cavity detuning.

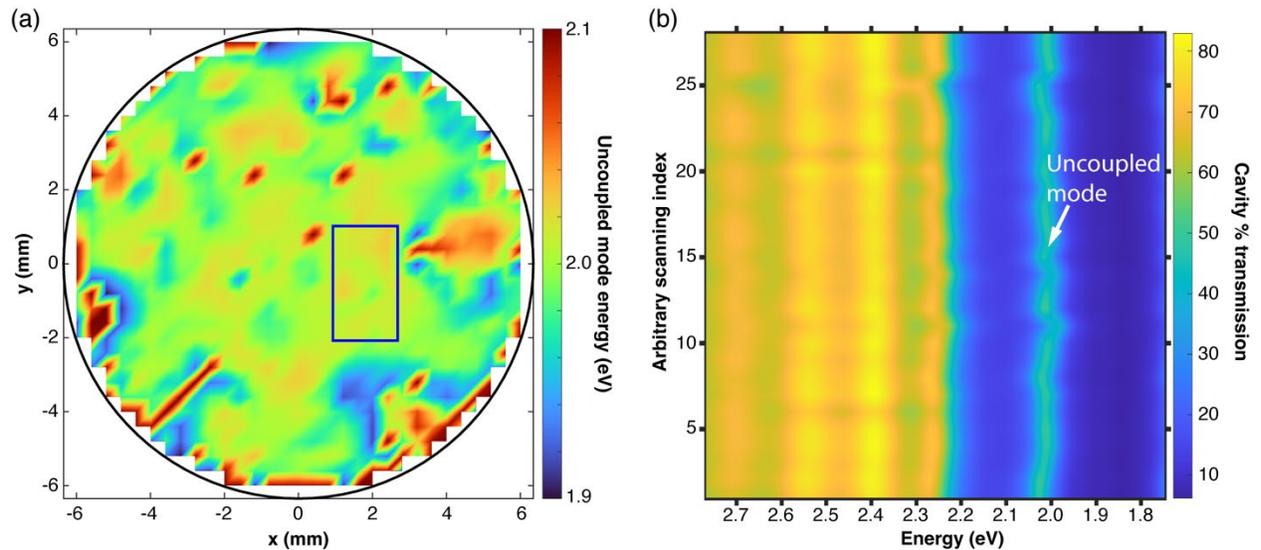

**Figure S5.** Spatial cavity-coupling map and pump-probe spectra of Cavity 3, which is filled with an 816 nm thick Ce6T/PS film and features a cavity mode that is red-detuned from resonance with the Ce6T $Q_y$ band. (a) Spatial cavity-coupling map tracking the energy of the cavity mode at 2.01 eV as a function of $x$ and $y$ coordinates in the cavity plane. The uniform region used for rastering during pump-probe experiments is highlighted in the blue box. (b) Relatively consistent cavity transmission spectra are acquired while rastering over the boxed region of the map shown in panel (a), indicating that rastering can be performed over this region without significant cavity detuning.

## S4. TCSPC and ultrafast pump-probe data for extracavity Ce6T solutions

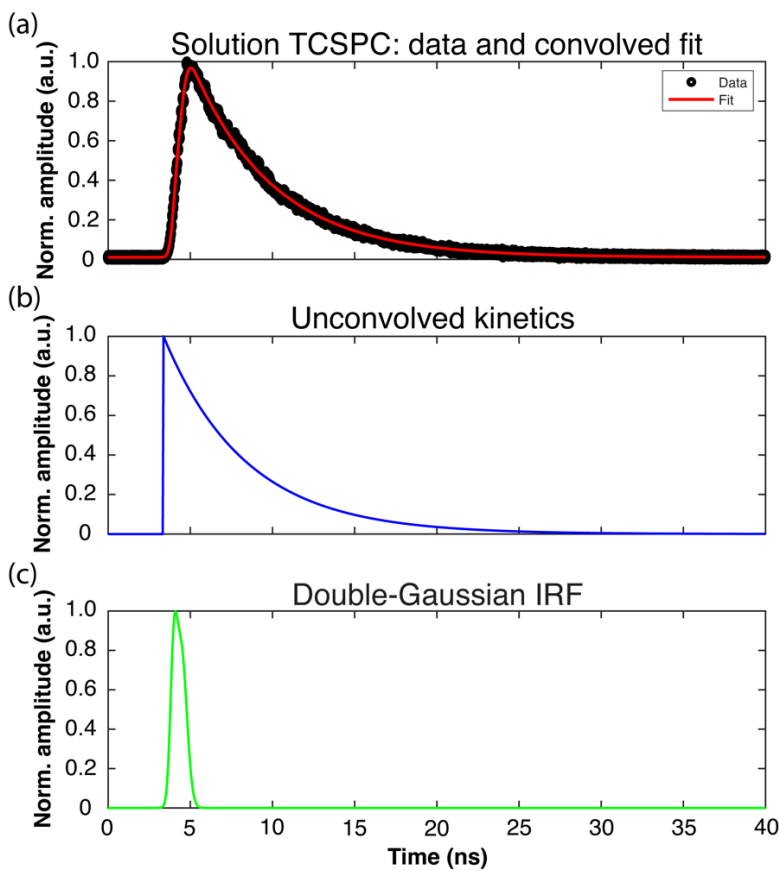

**Figure S6.** Ultrafast TCSPC emission data of an extracavity dilute solution of 10 μM Ce6T in toluene. Data is collected using a photoluminescence spectrometer (FLS980, Edinburgh Instruments) equipped with a picosecond pulsed diode laser (EPL-405, Edinburgh Instruments) which excites the sample at 3.09 eV (401 nm) with a pulse width of 58.1 ps. Fitting is performed using the lsqcurvefit function in MATLAB. (a) Ultrafast emission data (black dots) acquired with excitation at 3.09 eV and probing of the monomer Ce6T emission at 1.84 eV (673 nm). We fit these data (red line) to the convolution of a single exponential decay with the IRF. For these data, the IRF is obtained from the fit of the scattering signal in a solution of silica nanoparticles, as described above in Section S2. (b) The fitted exponential decay kinetics, which feature a decay time constant of 4.974 ± 0.011 ns. (c) Experimental IRF reproduced from Fig. S3.

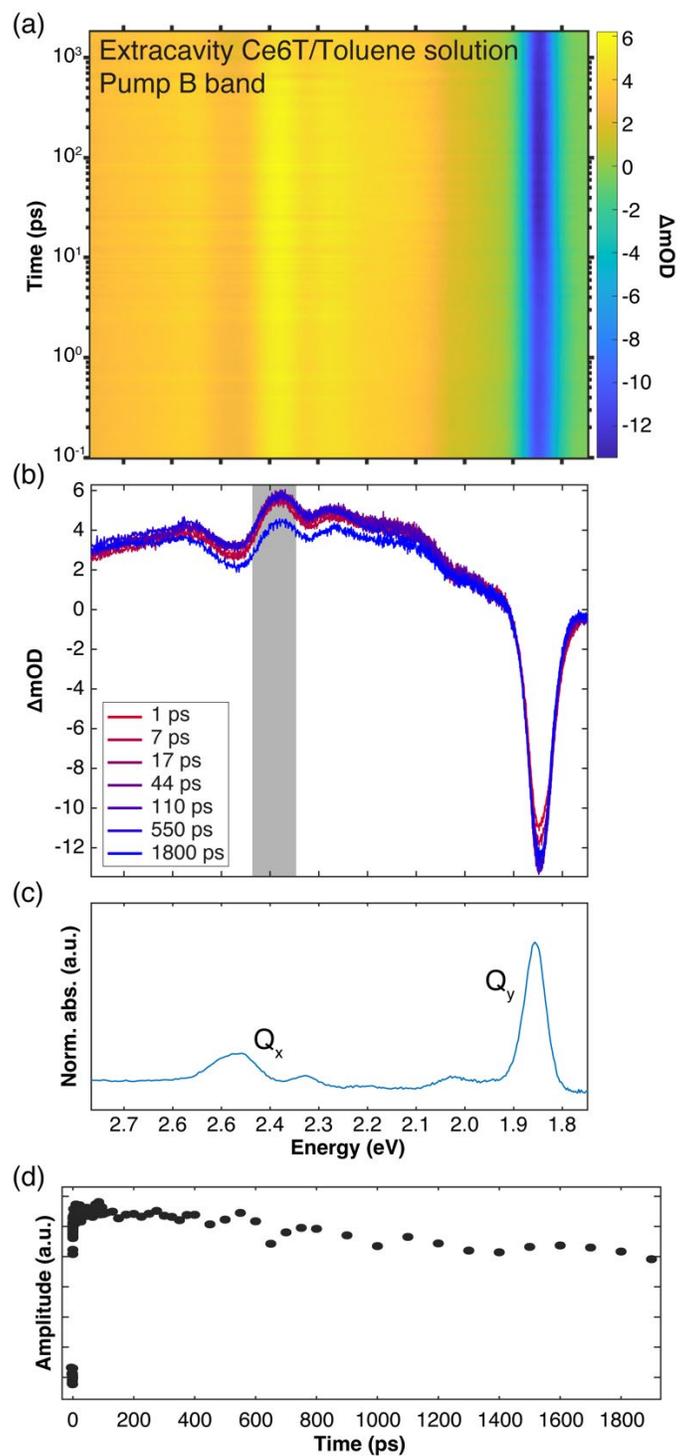

**Figure S7.** Transient dynamics of an extracavity solution of 100 µM Ce6T in toluene following optical excitation of the B band at 3.10 eV (400 nm). (a) Broadband pump-probe spectra and (b) representative spectral linecuts. (c) Linear absorption spectrum of the same Ce6T/toluene solution to illustrate where relevant spectral features lie. (d) Temporal linecut of the pump-probe data from panel (a) showing the ESA dynamics averaged over the spectral window from 2.35–2.42 eV (as marked in gray in panel (b)). The ESA feature is considerably longer-lived in this dilute solution than it is in dense thin films, as has been observed in prior literature, and as is also evident from the TCSPC data shown above in Fig. S6.[1–4]

## S5. Ultrafast pump-probe spectra for extracavity Ce6T/PS thin films pumping the $Q_y$ band

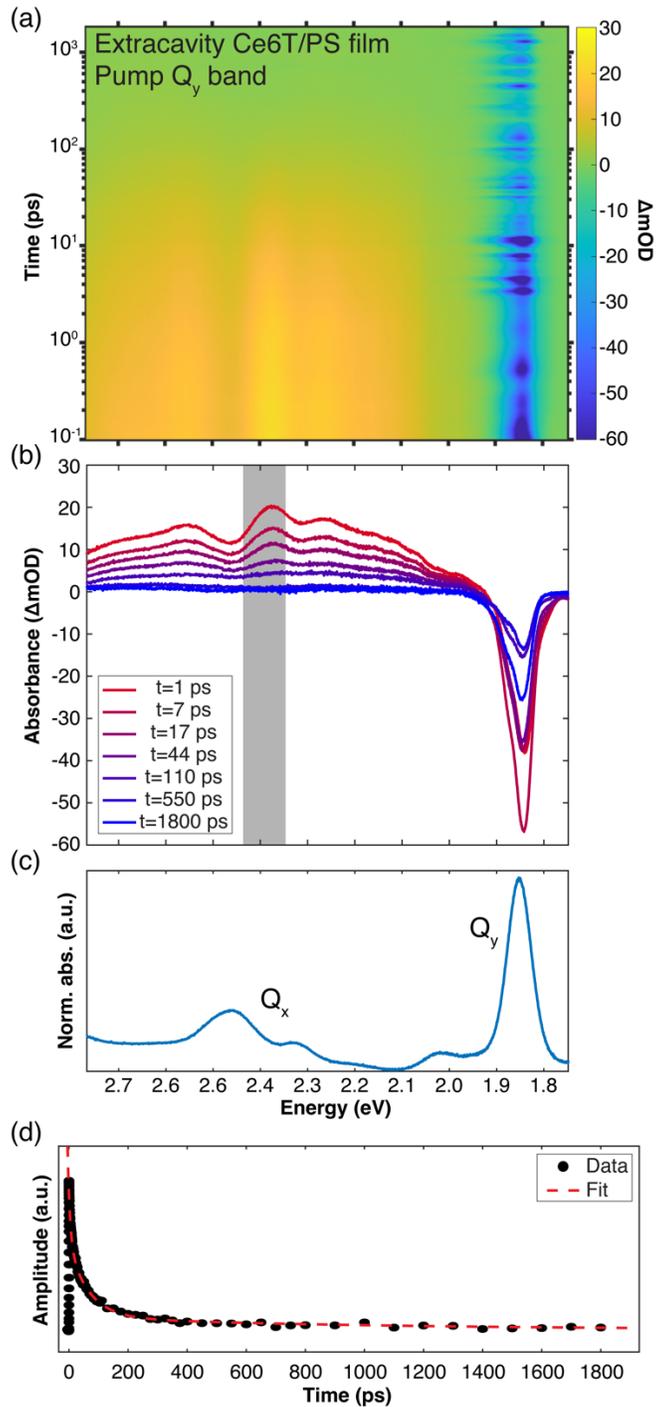

**Figure S8.** Transient dynamics of an extracavity Ce6T/PS film following optical excitation of the $Q_y$ band at 1.85 eV (669 nm). (a) Broadband pump-probe spectra and (b) representative spectral linecuts. Some noise appears at 1.85 eV due to pump scatter. (c) Linear absorption spectrum of Ce6T/PS replotted from Fig. 4a to illustrate where relevant spectral features lie. (d) Temporal linecut of the pump-probe data from panel (a) showing the ESA dynamics averaged over the spectral window from 2.35–2.42 eV (as marked in gray in panel (b)). Experimental data points are shown with black dots, while the red dashed line represents a fit of these data to three parallel exponential decays.

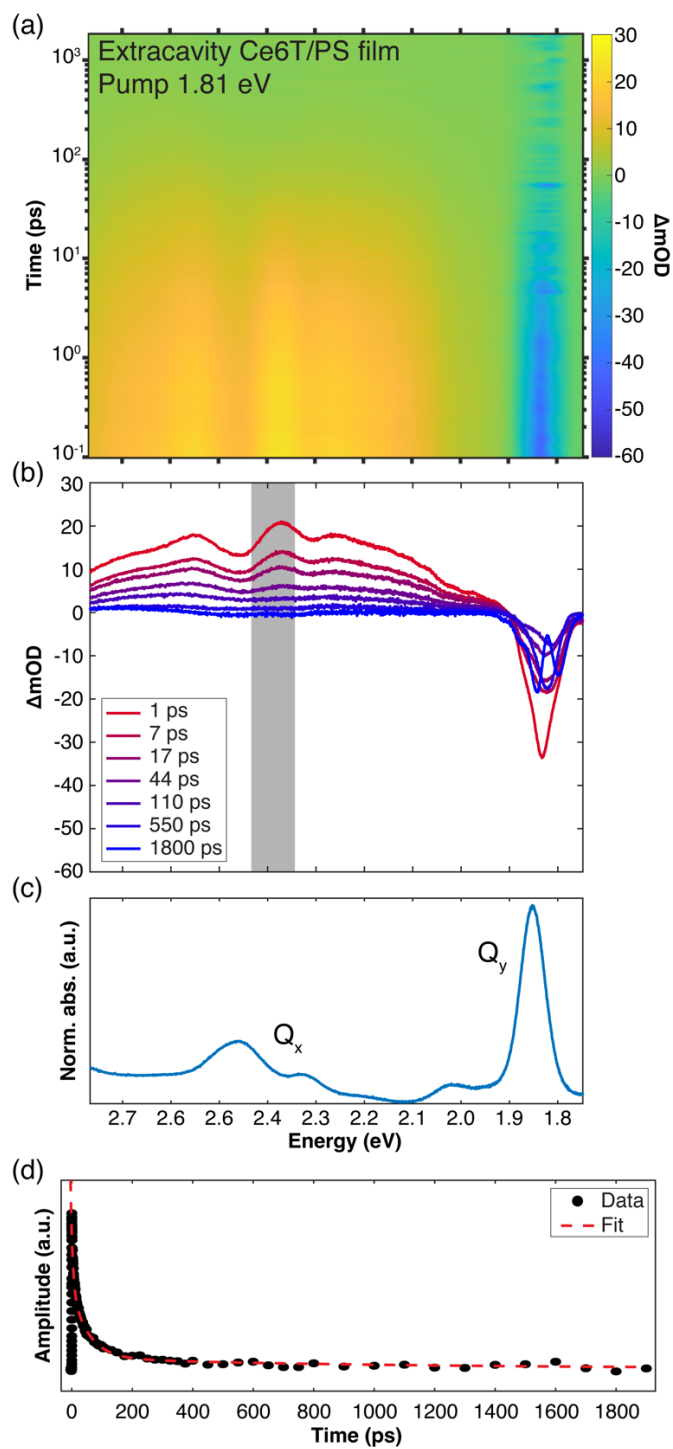

**Figure S9.** Transient dynamics of an extracavity Ce6T/PS film following optical excitation red of the $Q_y$ band at 1.81 eV (684 nm). This represents an extracavity control experiment for excitation of the lower polariton in intracavity experiments. (a) Broadband pump-probe spectra and (b) representative spectral linecuts. Some noise appears at 1.81 eV due to pump scatter. (c) Linear absorption spectrum of Ce6T/PS replotted from Fig. 4a to illustrate where relevant spectral features lie. (d) Temporal linecut of the pump-probe data from panel (a) showing the ESA dynamics averaged over the spectral window from 2.35–2.42 eV (as marked in gray in panel (b)). Experimental data points are shown with black dots, while the red dashed line represents a fit of these data to three parallel exponential decays.

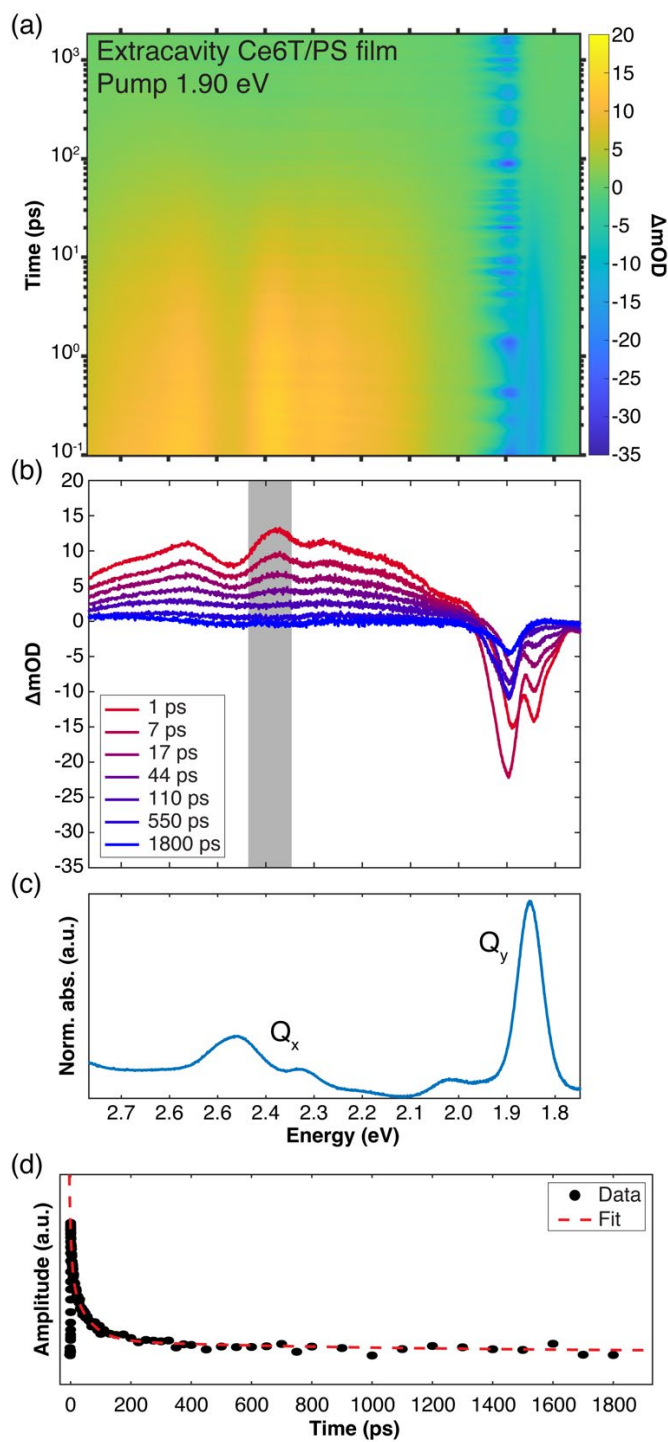

**Figure S10.** Transient dynamics of an extracavity Ce6T/PS film following optical excitation blue of the $Q_y$ band at 1.90 eV (653 nm). This represents an extracavity control experiment for excitation of the upper polariton in intracavity experiments. (a) Broadband pump-probe spectra and (b) representative spectral linecuts. Some noise appears at 1.90 eV due to pump scatter. (c) Linear absorption spectrum of Ce6T/PS replotted from Fig. 4a to illustrate where relevant spectral features lie. (d) Temporal linecut of the pump-probe data from panel (a) showing the ESA dynamics averaged over the spectral window from 2.35–2.42 eV (as marked in gray in panel (b)). Experimental data points are shown with black dots, while the red dashed line represents a fit of these data to three parallel exponential decays.

## S6. TCSPC data and ultrafast pump-probe spectra of excimer stimulated emission (SE) for extracavity Ce6T/PS thin films

We acquire TCSPC traces for six extracavity Ce6T/PS films, exciting the sample at 3.09 eV (401 nm) and probing at 1.68 eV (740 nm) to characterize excimer emission dynamics. Data from one representative film is shown in Fig. S11 below, wherein we fit the TCSPC data to the convolution of the IRF and two parallel decaying exponentials. The thin film TCSPC data kinetics are much faster than those of solution-phase Ce6T. As a result, we find that representing the IRF exactly via spline fits of the experimental silica nanoparticle scattering data (e.g. black dots in Fig. 3) yields better fits than representing the IRF with two fitted Gaussians (e.g. red trace in Fig. 3). We find decay time constants of 230 ± 30 ps and 1170 ± 120 ps for the excimer TCSPC signal in extracavity Ce6T/PS films. Our analysis of the excimer kinetics differs from that of Biswas et al.[5] who instead report a rise of 200 ps followed by a decay of 1.1 ns. When we fit our TCSPC data to a convolution of the experimental IRF with an exponential rise and two decays, we find a much shorter rise time constant of ~20 ps ± 20 ps that TCSPC does not have sufficient time resolution to capture. As a result, we do not include a rise in our TCSPC fits.

We find consistent results using ultrafast pump-probe spectroscopy to track stimulated emission (SE) signatures that report on the same excimer dynamics (see Fig. S12 below and Section III.B of the main text). We analyze the pump-probe SE data using similar practices as for analysis of the ESA dynamics. We take a temporal linecut of these data averaged over the spectral window from 1.67–1.72 eV. We trim this linecut before 1.5 ps and fit the longer-time dynamics to an exponential rise and two parallel decays (plus a constant offset) using the nonlinear least squares method in MATLAB. Results from individual fits to our data can be found in the accompanying Excel sheet. Fitting the SE data to an exponential rise and two parallel decays, we find an excimer rise time of 7.9 ± 0.5 ps, and decay time constants of 240 ± 30 ps and 1300 ± 110 ps, completely consistent with the TCSPC data. Our use of two independent spectroscopic techniques lends confidence that we are correctly capturing the Ce6T excimer rise and decay dynamics. In addition, our quoted ~8 ps excimer rise time is more consistent with the excimer formation time observed in other non-diffusion-limited systems.[6–9]

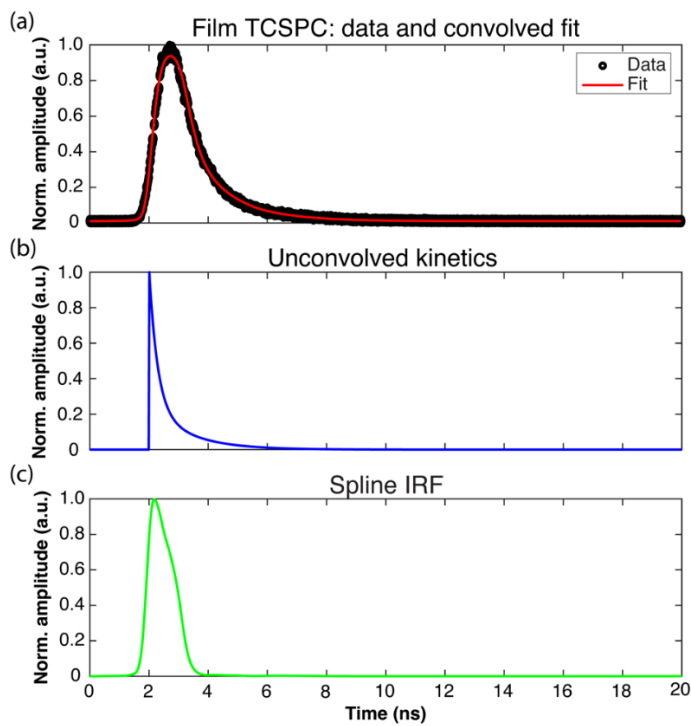

**Figure S11.** Ultrafast TCSPC emission data for a Ce6T/PS thin film. Data is collected using a photoluminescence spectrometer (FLS980, Edinburgh Instruments) fitted with a picosecond pulsed diode laser (EPL-405, Edinburgh Instruments) which excites the sample at 3.09 eV (401 nm) with a pulse width of 58.1 ps. Fitting is performed using the lsqcurvefit function in MATLAB. (a) Ultrafast emission data (black dots) acquired with 3.09 eV excitation and probing of the Ce6T excimer emission at 1.68 eV (740 nm). We fit (red line) these data to the convolution of two parallel exponents decays with the experimental IRF. For these data, the experimental IRF was represented by fitting a spline to the exact experimental IRF lineshape obtained from the scattering signal of a silica nanoparticle solution. This was done to capture smaller transient features in the scattering signal that bias the shorter emission decay in thin films. (b) The extracted Ce6T TCSPC kinetics after fitting. (c) The experimental IRF acquired by averaging together three spline fits of silica nanoparticle scattering data.

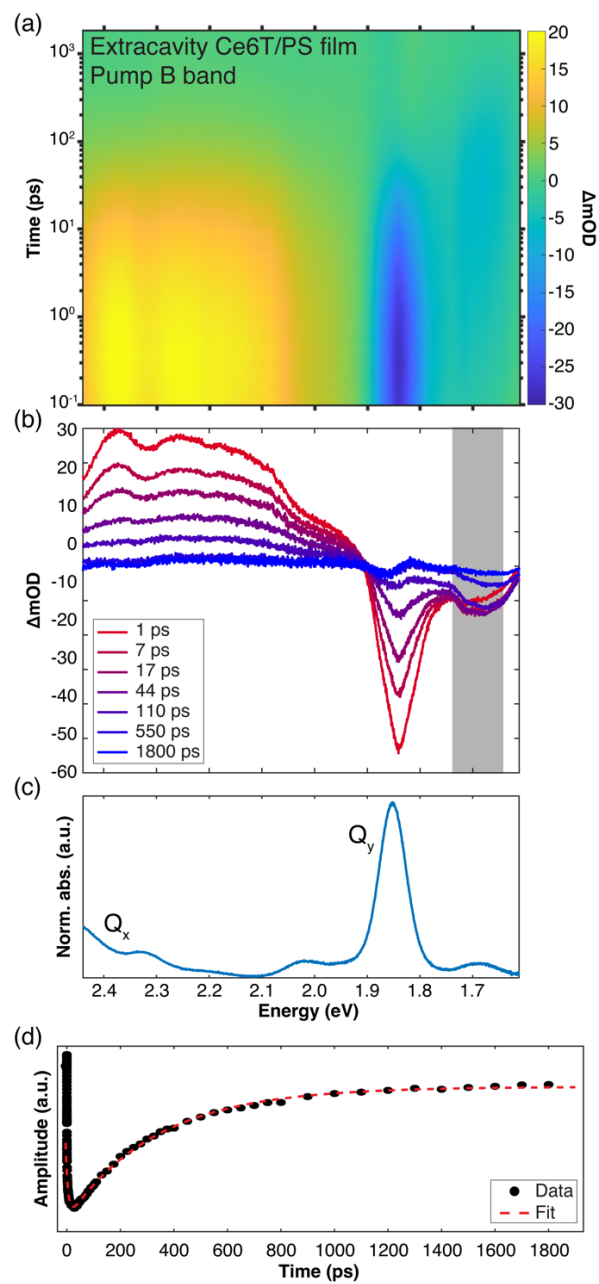

**Figure S12.** Transient dynamics of an extracavity Ce6T/PS film following optical excitation of the B band at 3.10 eV (400 nm), with an emphasis on the signatures of excimer stimulated emission (SE) near 1.7 eV. (a) Broadband pump-probe spectra and (b) representative spectral linecuts. (c) Linear absorption spectrum of Ce6T/PS replotted from Fig. 4a to illustrate where relevant spectral features lie. (d) Temporal linecut of the pump-probe data from panel (a) showing the SE dynamics averaged over the spectral window from 1.67– 1.72 eV (as marked in gray in panel (b)). Experimental data points are shown with black dots, while the red dashed line represents a fit of these data to an exponential rise and two parallel exponential decays.
S11

## S7. Diagrammatic representation of Ce6T/PS thin film excited state dynamics

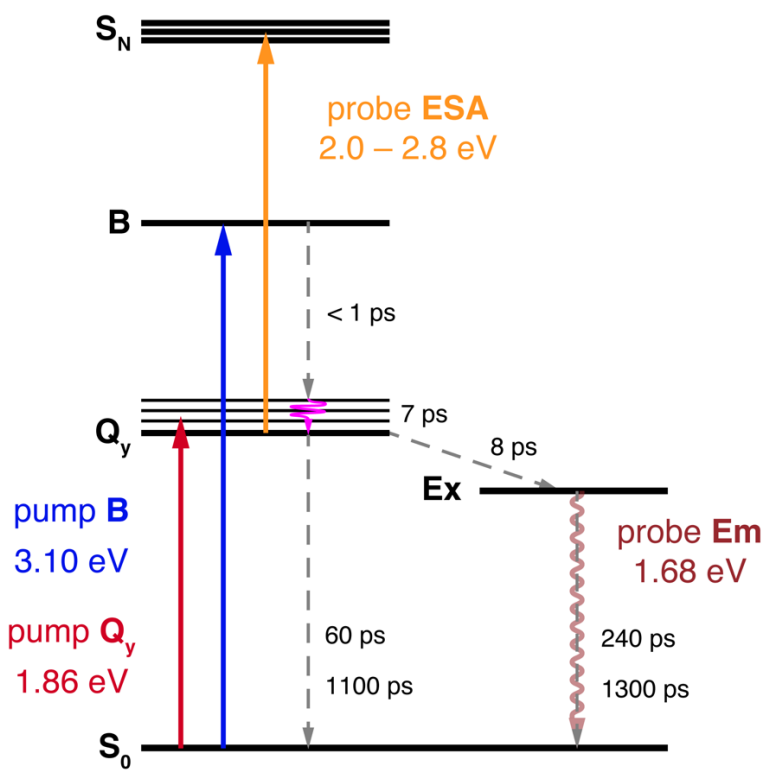

**Figure S13.** Energy diagram for Ce6T dynamics observed in PS thin films. Reported relaxation time constants are acquired from excited state absorption (ESA) and emission (Em) signatures from pump-probe and TCSPC experiments. Solid straight lines indicate optical excitation, solid wavy lines indicate excited state vibrational cooling or fluorescence readout, and dashed lines indicate nonradiative relaxation.

## S8. Ultrafast pump-probe spectra of Cavities 2 and 3 pumping the B band at 3.10 eV

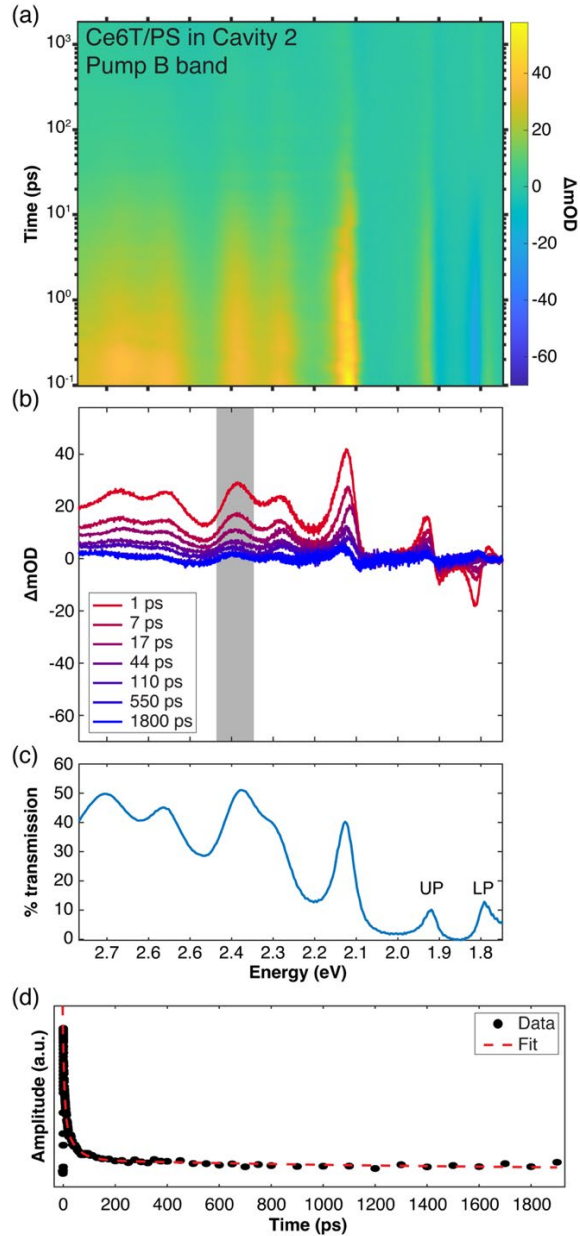

**Figure S14.** Transient dynamics of a Ce6T/PS film under strong coupling of the $Q_y$ band in DBR Cavity 2 following optical excitation of the B band at 3.10 eV (400 nm). These data are nearly identical to those acquired for Cavity 1 shown in Fig. 8 of the main text. (a) Broadband pump-probe spectra and (b) representative spectral linecuts. Excited state absorption (ESA) features are clearly visible from 2.2–2.8 eV through the transparent region of the DBR mirrors. Two derivative-like features appear at the energies of the polaritons on either side of the $Q_y$ band. These arise due to a contraction of the Rabi splitting from bleaching of the $S_0 \rightarrow Q_y$ transition. An additional derivative-like lineshape at 2.1 eV is representative of an uncoupled cavity mode shifting in energy due to modulation of the intracavity background refractive index. The $Q_y$ bleach feature at 1.86 eV is missing in these data compared to the pump-probe spectra in Cavity 1 where it is readily apparent. This is likely due to the larger Rabi splitting of Cavity 2 which places the polariton transmission windows of the cavity further from the $Q_y$ bleach, filtering any signature of the bleach out with greater coupling strength. (c) Linear transmission spectrum of Cavity 2 replotted from Fig. 7c of the main text. (d) Temporal linecut of the pump-probe data from panel (a) showing the ESA decay averaged over the spectral window from 2.35–2.42 eV (as marked in gray in panel (b)). Experimental data points are shown with black dots, while the red dashed line represents a fit of these data to three parallel exponential decays.

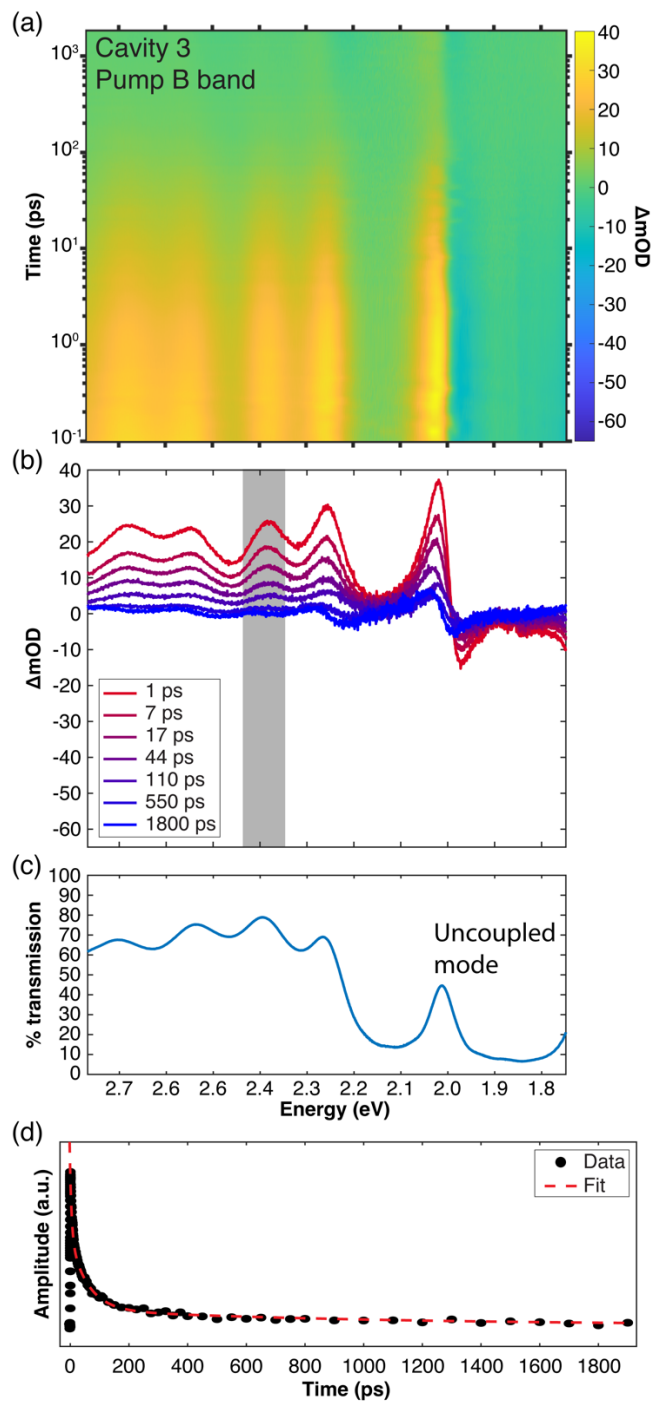

**Figure S15.** Transient dynamics of a Ce6T/PS film in Cavity 3 following optical excitation of the B band at 3.10 eV (400 nm). (a) Broadband pump-probe spectra and (b) representative spectral linecuts. Excited state absorption (ESA) features are clearly visible from 2.2−2.8 eV through the transparent region of the DBRs. (c) Linear transmission spectrum of Cavity 3, showing the off-resonance cavity mode near 2.01 eV. (d) Temporal linecut of the pump-probe data from panel (a) showing the ESA decay averaged over the spectral window from 2.35–2.42 eV (as marked in gray in panel (b)). Experimental data points are shown with black dots, while the red dashed line represents a fit of these data to three parallel exponential decays.

S14

## S9. Ultrafast pump-probe spectra of Cavity 1 pumping the lower and upper polaritons at 1.81 eV and 1.90 eV

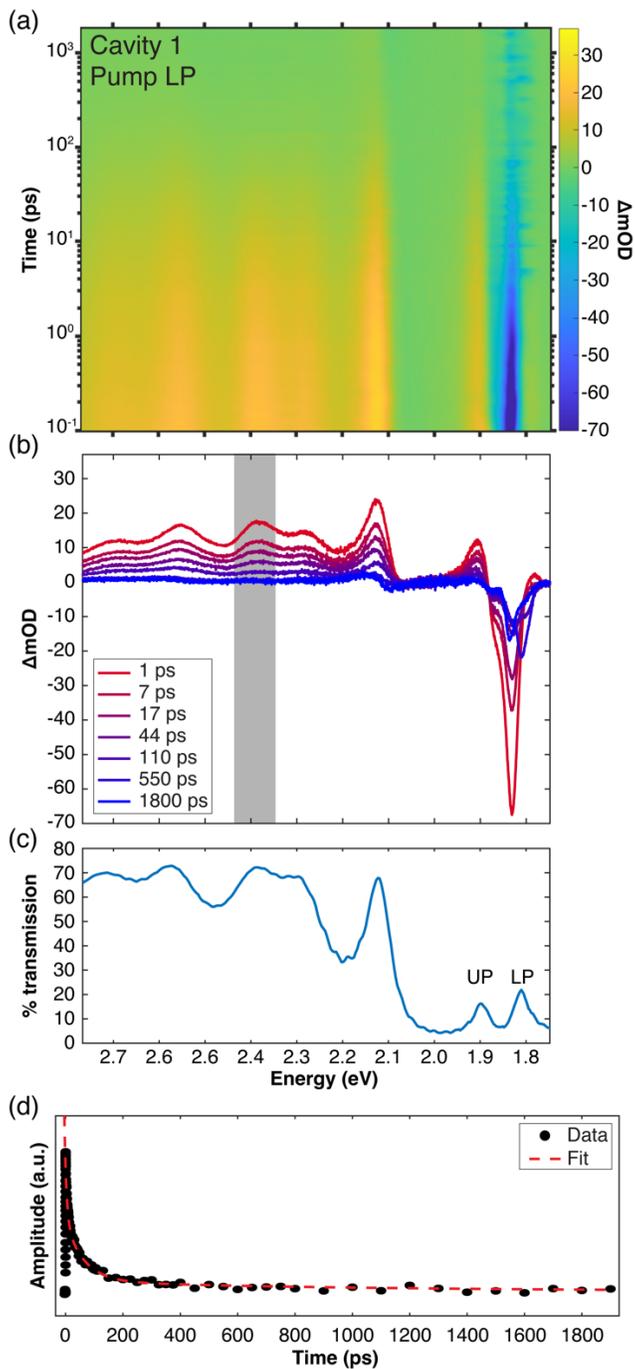

**Figure S16.** Transient dynamics of a Ce6T/PS film under strong coupling of the $Q_y$ band in DBR Cavity 1 following optical excitation of the lower polariton (LP) at 1.81 eV (684 nm). (a) Broadband pump-probe spectra and (b) representative spectral linecuts. Excited state absorption (ESA) features are visible from 2.2−2.8 eV through the transparent region of the DBR mirrors. (c) Linear transmission spectrum of Cavity 1 replotted from Fig. 7a of the main text. (d) Temporal linecut of the pump-probe data from panel (a) showing the ESA decay averaged over the spectral window from 2.35–2.42 eV (as marked in gray in panel (b)). Experimental data points are shown with black dots, while the red dashed line represents a fit of these data to three parallel exponential decays.

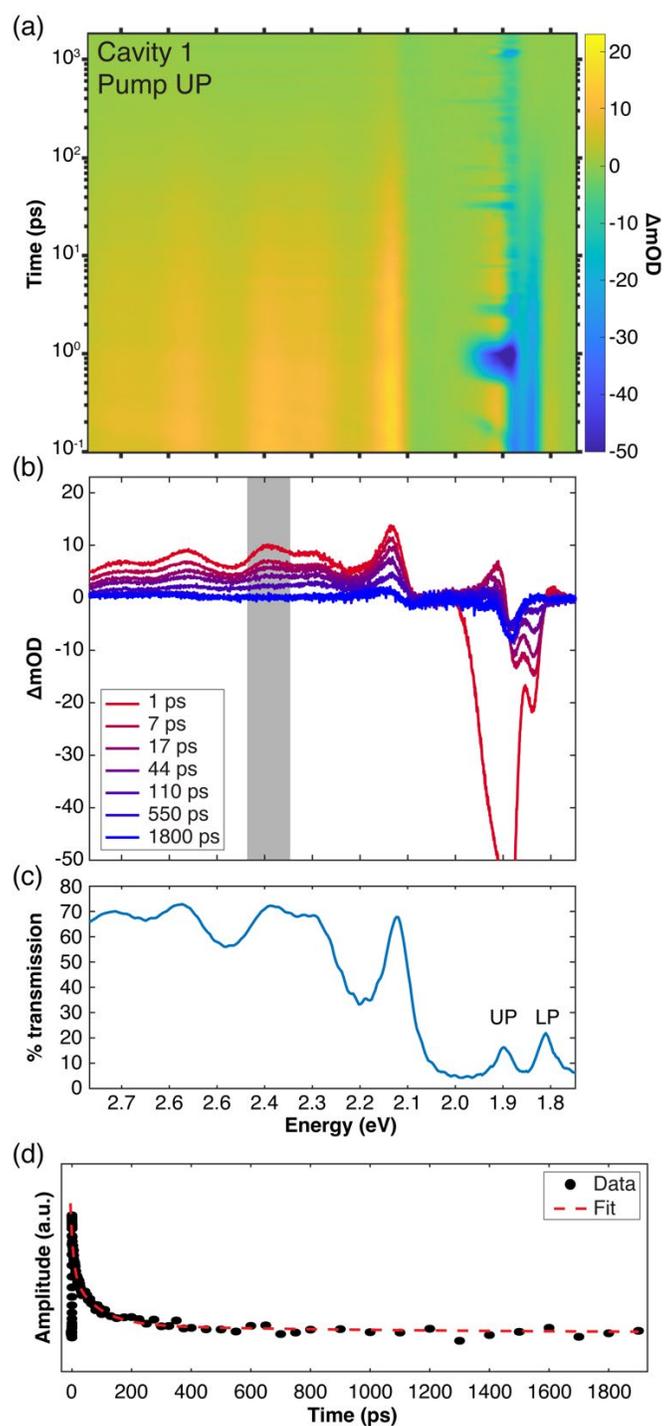

**Figure S17.** Transient dynamics of a Ce6T/PS film under strong coupling of the $Q_y$ band in DBR Cavity 1 following optical excitation of the upper polariton (UP) at 1.90 eV. **(a)** Broadband pump-probe spectra and **(b)** representative spectral linecuts. Excited state absorption (ESA) features are visible from 2.2−2.8 eV through the transparent region of the DBR mirrors. **(c)** Linear transmission spectrum of Cavity 1 replotted from Fig. 7a of the main text. (d) Temporal linecut of the pump-probe data from panel (a) showing the ESA decay averaged over the spectral window from 2.35–2.42 eV (as marked in gray in panel (b)). Experimental data points are shown with black dots, while the red dashed line represents a fit of these data to three parallel exponential decays.

# S10. Ultrafast response of the UV fused silica substrate, a single DBR mirror, a PS thin film, and an empty cavity

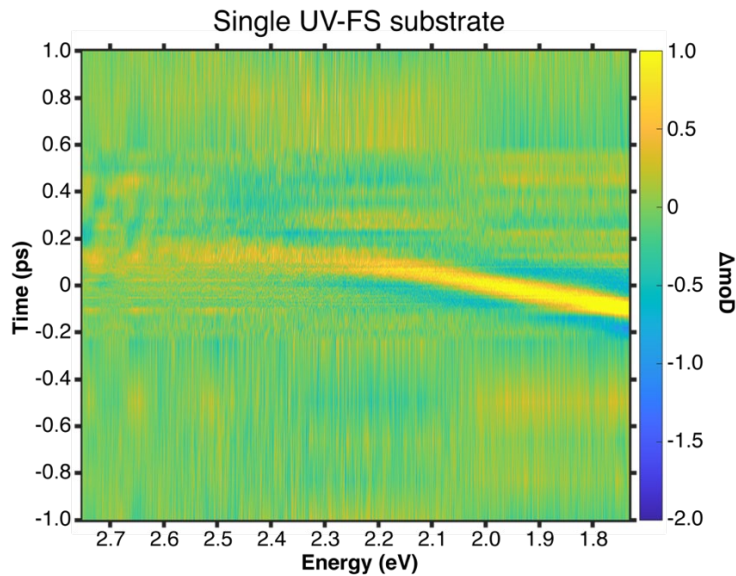

**Figure S18.** Transient pump-probe spectrum of a single 3 mm thick UV-fused silica (UV-FS) substrate acquired with a 3.10 eV pump. These data show no significant transient signals except for a pump-probe overlap feature near time zero. These data are not chirp corrected.

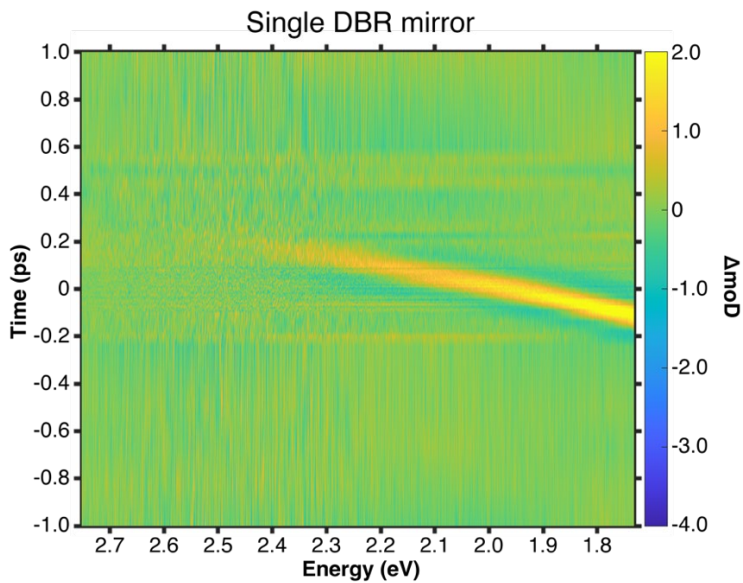

**Figure S19.** Transient pump-probe spectrum of a single DBR mirror acquired with a 3.10 eV pump pulse. These data show no significant transient signals except for a pump-probe overlap feature near time zero. These data are not chirp corrected.

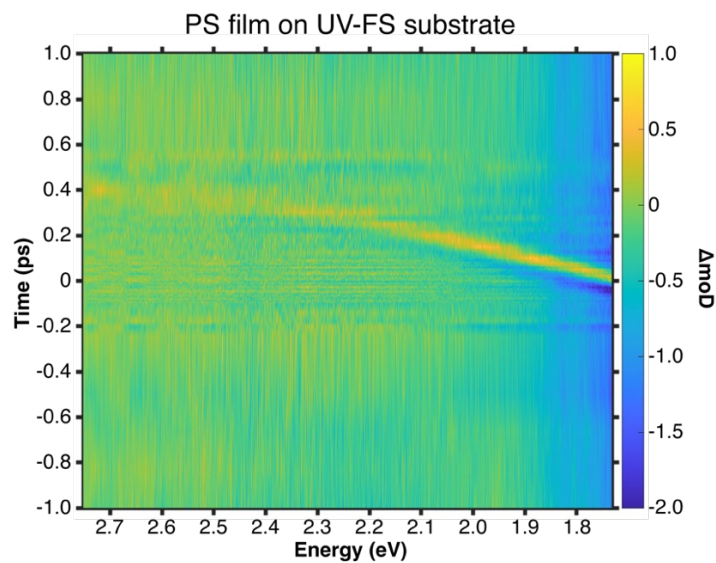

**Figure S20.** Transient pump-probe spectrum of a bare 705 nm polystyrene (PS) film on UV-FS acquired with a 3.10 eV pump. These data show no significant transient signals except for a pump-probe overlap feature near time zero. These data are not chirp corrected.

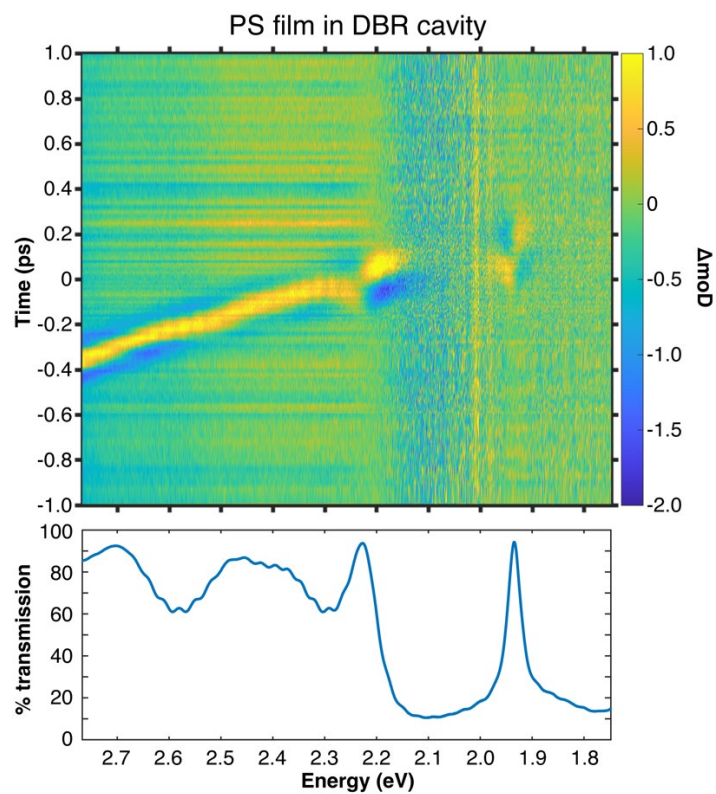

**Figure S21.** Transient pump-probe spectrum of a two-mirror DBR cavity containing just a 799 nm polystyrene (PS) film, acquired with a 3.10 eV pump. These data show no significant signals except for a short overlap feature near time zero. These data are not chirp corrected. The noise evident at probe energies below 1.9 eV and from 2.0–2.2 eV is due to low transmission of light through the cavity at these energies. The lower panel plots the linear cavity transmission spectrum to illustrate the position of a cavity fringe in the DBR mirror stop band at ~1.93 eV.

S18